\documentclass[10pt,twocolumn,letterpaper]{article}

\usepackage{iccv}              % To produce the CAMERA-READY version
\usepackage{soul}
\usepackage{bold-extra}
\usepackage{multirow}
\usepackage[inline]{enumitem}

\newcommand{\papertitle}{{\upshape{\bfseries\scshape SeHDR}}\xspace}

\newcommand{\bracketgs}{\textbf{Bracketed 3D Gaussians}\xspace}

\newcommand{\yy}[1]{\textcolor{black}{#1}}

\newcommand{\why}[1]{\textcolor{black}{#1}}

\newcommand{\ryn}{\textcolor[rgb]{0,0,0}}
\newcommand{\rynq}{\textcolor[rgb]{0,0,0}}

\definecolor{iccvblue}{rgb}{0.21,0.49,0.74}
\usepackage[pagebackref, breaklinks, colorlinks, allcolors=iccvblue]{hyperref}
\usepackage[accsupp]{axessibility}  % Improves PDF readability for those with disabilities.

\def\paperID{11685} % *** Enter the Paper ID here
\def\confName{ICCV}
\def\confYear{2025}

\title{\papertitle: Single-Exposure HDR Novel View Synthesis via 3D Gaussian Bracketing}

\author{Yiyu Li\qquad Haoyuan Wang\qquad Ke Xu\qquad Gerhard Petrus Hancke\qquad Rynson W.H. Lau\\
City University of Hong Kong\\
{\tt\small \{yiyuli.cs, cs.why\}@my.cityu.edu.hk, kkangwing@gmail.com}\\
{\tt\small \{gp.hancke,  Rynson.Lau\}@cityu.edu.hk}
}

\begin{document}

\twocolumn[{
\renewcommand\twocolumn[1][]{#1}
\maketitle
\begin{center}
    \captionsetup{type=figure}

    \vspace{-5mm}
    
    \includegraphics[width=\linewidth]{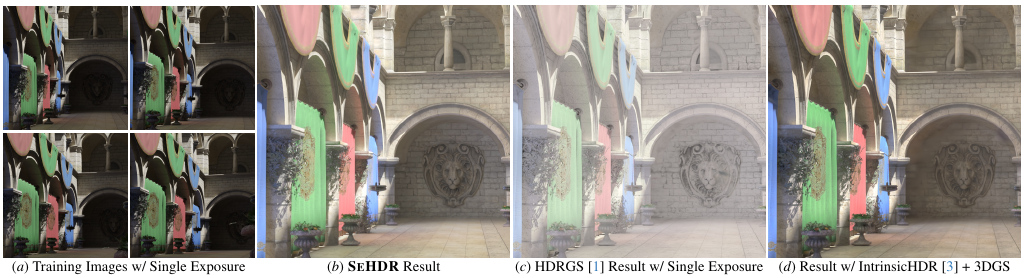}

    \vspace{-2mm}
    
    \captionof{figure}{Given a set of multi-view images with single exposure (\textit{a}), our method renders a photorealistic high quality novel view image with accurate color and fine details (\textit{b}), while existing multi-exposure based HDRGS~\cite{Cai_2024_nips_hdrgs} method fails to reconstruct HDR colors from a single-exposure (\textit{c}). Combining pre-trained HDR image reconstruction methods with 3DGS tends to fail to render accurate details and produce blurs and floaters,~\eg, on the stonewall region (\textit{d}). Note that both \papertitle and HDRGS are trained without HDR ground truth. We use PhotoMatix Pro~\cite{photomatixpro} \yy{with the same settings} to tone-map HDR results for visualization.}

    \vspace{1mm}

    \label{fig:teaser}
    
\end{center}

}]

\begin{abstract}

\vspace{-3mm}

This paper presents \papertitle, a novel high dynamic range 3D Gaussian Splatting (HDR-3DGS) approach for \ryn{generating} HDR novel views \ryn{given multi-view LDR} images.
Unlike existing methods that typically require the multi-view LDR input images \ryn{to be captured from different}
%\whydel{to have} \why{with} multiple 
exposures, which are tedious to capture and more likely to suffer from errors (\eg, object motion blurs and calibration/alignment inaccuracies),
our approach learns the HDR scene representation from multi-view LDR images of a single exposure. 
%\yy{Inspired by exposure-bracketing techniques in photography, which synthesize HDR images by merging multi-exposure LDR captures at a fixed camera pose, we adapt this concept to 3D scene representation.
%\papertitle estimates and fuses pseudo-multi-exposure information from single-exposure multi-view LDR inputs, thereby expanding the dynamic range of the HDR scene representation during optimization.}
\rynq{
%The challenge \yy{lies in compensating for quantization/saturation losses inherent to single-exposure inputs, which is essential to expand the dynamic range of 3DGS.}
%is to compensate for \yy{loss of quantized/saturated information inherent to} single exposure to expand the dynamic ranges of 3DGS during optimization.
%%if we can \yy{estimate} 
Our key insight to this ill-posed problem is that \ryn{by first}
\yy{estimating} \bracketgs (\ie, with different exposures) from single-exposure multi-view LDR \ryn{images,} %, we \yy{will} be able 
%\yy{enables us} 
\ryn{we may then be able} to merge these bracketed 3D Gaussians into \yy{an} HDR scene representation.}
%(*** This insight sounds a bit engineering, as it simply states what we do. My feeling is that our insight/observation should lead to answering the following question: why would we be able to expand multi-view LDR images of only a single exposure (instead of multiple exposures) into an HDR representation? Here, we only say that we will expand it into HDR during the optimization, which is not an insight. ***)
Specifically, \papertitle first learns base \why{3D Gaussians} from single-exposure LDR inputs, where the spherical harmonics parameterize colors in \ryn{a} linear color space.
We then \yy{estimate} multiple \why{3D Gaussians} with identical geometry but varying linear colors conditioned on exposure manipulations.
Finally, we propose the \yy{Differentiable} Neural Exposure Fusion (NeEF) to integrate the base and \yy{estimated} \why{3D Gaussians} into HDR \why{Gaussians} for novel view rendering.
Extensive experiments demonstrate that \papertitle outperforms existing methods \ryn{as well as} carefully designed baselines.
Codes are available at \url{https://github.com/yiyulics/SeHDR}.
\vspace{-4mm}
\end{abstract}    
\section{Introduction}
\label{sec:intro}

% \input{figsntabs/teaser}

%Consequently, there is increasing demand for HDR view synthesis to enhance visual fidelity and realism in rendered outputs.

High Dynamic Range Novel View Synthesis (HDR-NVS) \why{has recently been proposed}~\cite{huang_2022_cvpr_hdrnerf}
% \rynq{(*** It may be good to add a reference here of the paper that proposed this task. ***)}
to reconstruct and render an HDR scene \why{with} novel viewpoints given Low Dynamic Range (LDR) input images.
Compared to previous LDR-NVS methods, the HDR scene representation can capture a wider spectrum of illumination intensities (\eg, ranging from intensive sunlight to dark shadows), preserving finer details of regions that are over-exposed and under-exposed in \ryn{the} LDR images, thereby significantly enhancing the fidelity and realism of generated novel images.
%and more accurately representing real-world lighting conditions.

%In contrast to LDR images, which typically have data values ranging from 0 to 255, HDR images encompass a range from 0 to $+\infty$.
%In addition, while LDR images are quantized to integers, HDR images are nearly continuous due to the floating point representation.
%This extended range enables HDR imaging to capture a wider spectrum of illumination intensities and a smoother light and shadow gradations, thereby preserving finer details in both over-exposed and under-exposed regions and more accurately representing real-world lighting conditions.

\ryn{A key requirement of existing HDR-NVS methods~\cite{huang_2022_cvpr_hdrnerf, Cai_2024_nips_hdrgs} is the use of multi-view LDR images with multiple exposures as input}.
%Existing HDR-NVS methods~\cite{huang_2022_cvpr_hdrnerf, Cai_2024_nips_hdrgs} primarily require multi-view LDR input images with multiple exposures.
%
While optimizing the scene geometries via NeRF~\cite{mildenhall2020nerf} or 3DGS~\cite{kerbl_2023_tog_3Dgaussians}, multi-view aligned pixels with different exposures
%Aligned pixels across views typically exhibit varying exposures, 
enable their networks to compensate for the missing information \ryn{(due to quantization and saturation)} from other views.
% with different exposures.
However, in practice, capturing multi-view multi-exposure images remains challenging even with tripod stabilization and software-based exposure control.
This arises from the strict alignment requirements: \ie, multi-exposure images within the same view must be perfectly registered while \yy{Colmap}~\cite{schoenberger_2016_cvpr_sfm, schoenberger_2016_eccv_mvs} features across views with different exposures should also be aligned precisely.
Such constraints make the capture process highly susceptible to motion blur and  prone to calibration/alignment errors \yy{when pose estimation is needed}, hindering the real-world applicability of these multi-exposure-based techniques.
%, severely limiting the real-world applicability of HDR-NVS approaches
% \why{(why: as far as I know, misalignment does not affect these methods like HDR-NeRF significantly.)}
\yy{Directly applying multi-exposure HDR-NVS methods to single-exposure inputs is inherently ineffective, as these approaches rely on multi-exposure references from alternate viewpoints to reconstruct accurate HDR color information (\cf, \cref{fig:teaser}(\textit{c})).}

To address this limitation, \ryn{in this work, we aim to learn} the HDR scene representation for NVS from single-exposure multi-view LDR images.
%, significantly broadening the applicability of HDR-NVS methods.
\ryn{However, unlike} previous HDR-NVS methods that can rely on the complementary exposure information of aligned pixels from adjacent views, we \ryn{need} to develop effective priors to constrain the \ryn{ill-posed nature of this problem,} based on single-exposure multi-view LDR input images.
%to compensate for the quantization and saturation \why{(can we?)} losses in single-exposure multi-view LDR input images.
A straightforward solution is to combine single image-based HDR reconstruction methods~\cite{endo_2017_tog_drtmo, Liu_2020_CVPR_singlehdr, dille_2024_eccv_IntrinsicHDR, Eilertsen_17_tog_hdrcnn, lee_2018_eccv_deepreversehdri} with 3DGS. However, such a strategy may not work well as those single-image HDR methods may introduce additional multi-view inconsistencies, resulting in obvious floater artifacts in the rendered novel view images \yy{(\cf, \cref{fig:teaser}(\textit{d}))}.
%However, as previously noted, existing HDR-NVS approaches require multi-exposure inputs to identify and preserve quantization and saturation losses across adjacent views with differing exposures.
%In single-exposure scenarios, aligned pixels share identical exposure settings and nearly indistinguishable brightness levels, making such losses irretrievable from other views.
%Consequently, compensating for quantization and saturation losses inherent to single-exposure inputs, which are critical for expanding the dynamic range of 3DGS~\cite{kerbl_2023_tog_3Dgaussians}, poses a significant challenge.

In this \ryn{paper}, we propose \papertitle, the first HDR-NVS framework capable of synthesizing HDR novel view images from single-exposure multi-view LDR inputs.
Inspired by exposure-bracketing techniques in computational photography, which either capture multi-exposure LDR images at a fixed camera pose~\cite{debevec_1997_siggraph_hdr, mertens_2007_cga_exposure_fusion} or synthesize multi-exposures from a single-exposure LDR input~\cite{endo_2017_tog_drtmo}, we propose to synthesize and then merge the \bracketgs 
%(\ie, sharing identical geometric properties such as spatial coordinates, Gaussian means, and covariance matrices but having varying exposures) into \ryn{an} HDR scene representation. 
\rynq{into \ryn{an} HDR scene representation. An important feature of these \bracketgs is that they share identical geometric properties, such as spatial coordinates, Gaussian means, and covariance matrices, but have varying exposures.}
% (*** See if this rewrite is OK to you. ***)}
%adapt these paradigms to merge bracketed 3D scenes into HDR scene representation.
% Our key insight is that estimating {\bf exposure-bracketed 3D Gaussians} with different exposures from single-exposure multi-view LDR inputs enables us to merge these bracketed 3D Gaussians into an HDR scene representation.
%Specifically, we present \bracketgs, share identical geometric properties (\ie, spatial coordinates, Gaussian means, and covariance matrices) but exhibit varying exposures.
Specifically, given single-exposure LDR inputs, we first learn a base set of 3D Gaussians whose spherical harmonics (SH) coefficients parameterize colors in \yy{an estimated} linearized radiance space.
By manipulating these linearized color values, we \why{then} estimate additional 3D Gaussians, conditioned on sampled exposures different from the exposure of \ryn{the} input LDR images, to \rynq{produce} bracketed 3D Gaussians.
Finally, we integrate the base and estimated Gaussians into \ryn{an} HDR scene representation via our proposed \yy{Differentiable} Neural Exposure Fusion (NeEF) method, \ryn{which operates} within the \ryn{spherical harmonics (SH)} parameter space.
Experimental results demonstrate that \papertitle 
\ryn{achieves superior visual fidelity, outperforming \ryn{SOTA} methods and carefully-designed baselines by 14.3dB without supervision on HDR results.}
%outperforms existing state-of-the-art methods and baselines \TODO{by xxxdB}, achieving superior visual fidelity.

\ryn{Our key} contributions can be summarized as:
\begin{itemize}
    \item We propose \papertitle, the first framework for High Dynamic Range Novel View Synthesis (HDR-NVS) based on 3D Gaussian Splatting, addressing the single-exposure HDR-NVS problem, a novel task to the best of our knowledge. \yy{In contrast to existing multi-exposure HDR-NVS methods, our approach is \ryn{applicable to any standard multi-view datasets seamlessly}, eliminating the need for specialized multi-exposure capture setups and thereby broadening practical applicability.}
    \item We introduce \bracketgs, extending exposure-bracketing principles to 3D space to synthesize multi-exposure 3D Gaussians for fusion into HDR 3D Gaussians. To enable this fusion, we further design a \yy{Differentiable} Neural Exposure Fusion (NeEF) method, \ryn{which operates within the SH space, to effectively reconstruct} HDR radiance fields.
    \item Extensive comparative experiments against state-of-the-art HDR-NVS methods and baselines demonstrate that our approach achieves superior \ryn{performances} in dynamic range expansion and photorealism.
\end{itemize}

\section{Related Works}
\label{sec:related-works}
\noindent \textbf{HDR-NVS.}
Recent advances have focused on directly reconstructing HDR scenes from multi-view LDR input images.
Huang~\etal~\cite{huang_2022_cvpr_hdrnerf} propose the first HDR-NVS method, HDR-NeRF, \ryn{which} integrates a simplified physical imaging process into the NeRF~\cite{mildenhall2020nerf} framework.
Unlike standard NeRF that models density and LDR color along a ray, HDR-NeRF encodes density and unbounded scene radiance (ranging from 0 to $+\infty$) along each ray, employing a learnable tone-mapping MLP to project radiance back into LDR RGB space.
\ryn{Recently, Cai~\etal~\cite{Cai_2024_nips_hdrgs} propose} to enhance the visual quality, training efficiency, and rendering speed by replacing the implicit 3D representation of NeRF with the explicit 3DGS and modeling scene radiance through SH of Gaussian point clouds.

However, both approaches require multi-exposure LDR inputs, which are labor-intensive to capture and prone to motion blur, calibration errors, and alignment inaccuracies.
To address these limitations, we propose \papertitle, a method for reconstructing HDR scenes directly from single-exposure LDR images, which is able to take any standard multi-view dataset as inputs.

\vspace{1ex}

\noindent \textbf{HDR Image Reconstruction.}
Conventional methods for reconstructing high dynamic range (HDR) images typically rely on two approaches: estimating an intermediate camera response function (CRF) from \ryn{multiple-exposure LDR} images~\cite{grossberg_2004_pami_emor, debevec_1997_siggraph_hdr}, or directly fusing bracketed LDR images with identical camera poses but varying exposures into HDR representations using intensity-prior-guided weight maps~\cite{mertens_2007_cga_exposure_fusion}.
These methods inherently require multiple exposure inputs to generate HDR outputs.
\ryn{The} advances in deep learning have enabled subsequent studies to explore single-exposure LDR-to-HDR reconstruction through \ryn{CNNs}~\cite{Eilertsen_17_tog_hdrcnn, endo_2017_tog_drtmo, Liu_2020_CVPR_singlehdr}, generative adversarial networks (GANs)~\cite{lee_2018_eccv_deepreversehdri}, and deep residual networks (ResNets)~\cite{dille_2024_eccv_IntrinsicHDR}.

\ryn{Although integrating} pre-trained single-image HDR reconstruction techniques with Gaussian splatting may be a plausible solution for the HDR-NVS problem,
\ryn{we show in this work that such a strategy often proves} ineffective due to multi-view inconsistencies introduced by single-image methods, resulting in obvious floater and blurry artifacts in the rendered novel view images (\cf, \cref{fig:teaser}(\textit{d})).

\vspace{1ex}

\noindent \textbf{Novel View Synthesis (NVS).}
Novel View Synthesis (NVS) aims to generate photorealistic images of a scene from unobserved viewpoints, given a set of posed input images.
State-of-the-art NVS methods typically rely on NeRF~\cite{mildenhall2020nerf} and 3DGS~\cite{kerbl_2023_tog_3Dgaussians} to parameterize the \ryn{scene} as implicit 3D position/views to color/density mappings and explicit 3D Gaussian points, respectively.
%to learn an implicit mapping function from the position of a 3D point and view direction to the point color and volume density, or based on 3D Gaussian Splatting (3DGS)~\cite{kerbl_2023_tog_3Dgaussians} to learn explicitly representations of a scene by millions of Gaussian point clouds.
While NVS has widespread applications in \eg, autonomous driving~\cite{yang_2020_cvpr_surfelgan, huang_2023_iccv_neural_lidar, zhou_2024_cvpr_drivinggaussian}, virtual reality~\cite{jiang_2024_siggraph_vrgs,GSS,dai2025_4dgv,dong2023sailor}, robotics~\cite{yan_2024_cvpr_gsslam, keetha_2024_cvpr_splatam, matsuki_2024_gsslam, li_2024_eccv_sgsslam, peng_2024_siggraph_rtgslam, hu_2024_eccv_cgslam}, scene understanding/editing~\cite{wang2023lighting,qu2024lush,zhou_2024_cvpr_feature3dgs, ye_2024_eccv_gaussian_grouping}, and physics simulation~\cite{zhong_2024_eccv_elastic_simulation, xie_2024_cvpr_physgaussian, zhang_2024_eccv_physdreamer},
it remains constrained by the low dynamic range of rendered images, which limits their ability to express a broad range of illumination.
This limitation may result in loss of detail in \ryn{over-exposed and under-exposed} regions, color shifts, and inability to replicate subtle light and shadow gradations, which are critical for achieving photorealism in human visual perception.

In this work, we aim to expand the dynamic ranges of 3DGS-based scene representations using single-exposure multi-view LDR images.

\section{Method}
\label{sec:methods}

\begin{figure*}[t]
    \centering
    \includegraphics[width=0.9\linewidth]{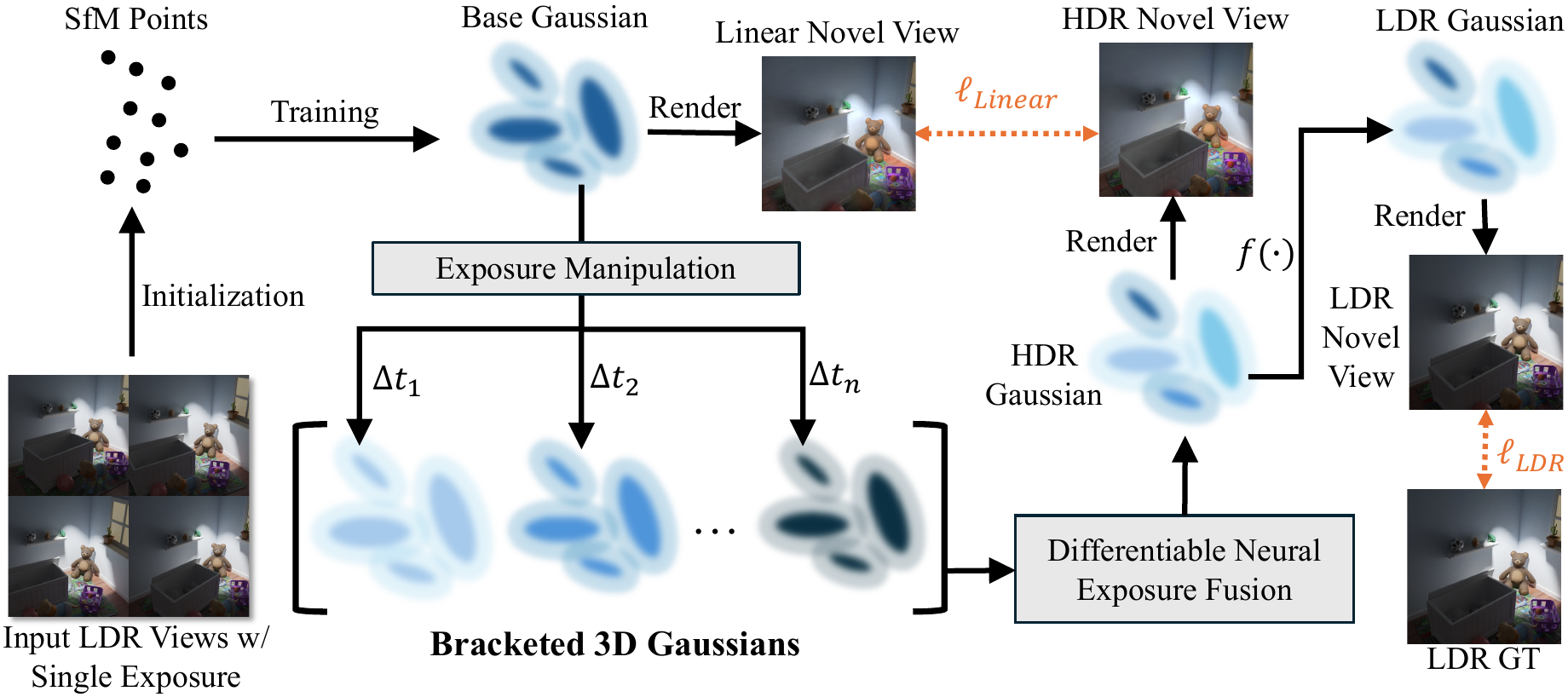}
    \vspace{-2mm}
    \caption{The pipeline of \papertitle. First, we leverage SfM~\cite{schoenberger_2016_cvpr_sfm} to refine camera parameters and initialize 3D Gaussian primitives. A base Gaussian representation is trained, where SH coefficients parameterize radiance in a linear color space aligned with the input exposure time. Second, we generate multi-exposure Gaussian variants (\bracketgs) by manipulating the linear radiance space with sampled exposure times, preserving shared geometry but with varying exposures. Finally, our proposed Differentiable Neural Exposure Fusion module fuses the bracketed Gaussians directly within the SH coefficient space, producing a unified HDR Gaussian representation capable of rendering photometrically consistent HDR novel views.}
    \vspace{-2mm}
    \label{fig:pipeline}
\end{figure*}

We propose \papertitle, a novel framework for reconstructing \yy{HDR 3D Gaussians} and rendering \yy{HDR novel views}
%\rynq{in  novel views HDR 3D Gaussians (*** This phrase reads a bit strange. Please just check to make sure that it is written correctly. ***)} 
from multi-view LDR images captured under a single exposure.
%, enabling HDR-NVS.
As illustrated in \cref{fig:pipeline}, our pipeline comprises three key stages.
First, we learn a base 3D Gaussian whose SH coefficients parameterize colors in a linear radiance space aligned with the exposure time \why{of the input images}. 
\ryn{Second}, we estimate \yy{multi-exposure Gaussian variants with shared geometry but varying exposures}, termed \bracketgs, by modulating the linear radiance space with sampled exposure times.
\ryn{Third}, we fuse the bracketed Gaussians into a unified HDR Gaussian representation using our proposed Differentiable Neural Exposure Fusion (NeEF) module, operating directly on the SH coefficient space.
This fused representation enables rendering of HDR novel views.

In this section, we will detail the technical foundation of our method: we first review preliminaries, formalize the bracketed Gaussians derivation, and then introduce the NeEF module.

\subsection{Preliminaries}

\noindent \textbf{Exposure Bracketing.}
A widely adopted image acquisition process~\cite{debevec_1997_siggraph_hdr} maps scene irradiance to pixel values through the following equation:
\begin{align}
    Z = f(E\Delta t),
    \label{eq:image-formation}
\end{align}
where $E$ represents scene irradiance (the HDR quantity in our framework), defined as the amount of light incident on the camera sensor.
$Z$ represents the pixel values (LDR outputs) generated by the acquisition process.
The exposure time $\Delta t$, governed by shutter speed, determines the brightness of the LDR image.
$f(\cdot)$ is the \textit{camera response function} (CRF)~\cite{dufaux2016high}, a nonlinear mapping \ryn{to compress} values from the \why{original} domain $(0, +\infty)$ to the normalized range $(0, 1)$. 

In computational photography, a set of images captured under identical camera poses and static scene radiance (\ie, $E$ remains constant) but with varying exposure times $\Delta t$ is called \textbf{\textit{bracketed images}}, denoted as $\{Z_1, ...Z_n\}$, where $n$ represents the number of exposures.
Given bracketed images $\{Z_1, ...Z_n\}$ and their corresponding exposure times $\{\Delta t_1, ...\Delta t_n\}$, reconstructing the corresponding HDR image becomes a well-posed task and the scene irradiance $E$ can be estimated accurately using the well-established HDR image reconstruction methods~\cite{debevec_1997_siggraph_hdr, mertens_2007_cga_exposure_fusion}.
%In contrast, our problem setup involves only a single exposure $Z$ and $\Delta t$, rendering $E$ underdetermined and resulting in an ill-posed inverse problem.
%\yycomment{(Optional) While multi-view images captured under a single exposure provide multiple observations of the scene, the corresponding pixel values for the same spatial point across different viewpoints exhibit minimal variation.
%Consequently, these multi-view inputs can be considered functionally equivalent to a single image with expanded spatial sampling, merely providing additional pixel values without introducing supplementary exposure-related information.
%This characteristic fails to mitigate the inherent ill-posedness of the inverse problem.}

\vspace{1ex}

\noindent \textbf{3D Gaussian Splatting.} 
3DGS~\cite{kerbl_2023_tog_3Dgaussians} models a scene using a collection of explicit 3D Gaussian primitives $\mathcal{G}$ defined as:
\vspace{-3mm}
\begin{align}
    \mathcal{G} = \{G_i (\mu_i, \Sigma_i, \alpha_i, c_i)|i=1, 2, \dots, N\},
\end{align}
where $N$ denotes the total number of Gaussians. Each primitive $G_i$ is parameterized by: \why{center coordinates $\mu_i\in \mathbb{R}^3$ in world space}, 3D covariance matrix $\Sigma_i\in \mathbb{R}^{3\times 3}$ defining the spatial scaling $S_i$ and rotation $R_i$ by
\begin{align}
    \Sigma_i = R_iS_iS_i^TR_i^T,
\end{align}
opacity $\alpha_i \in \mathbb{R}$ controlling transparency during rendering, and color $c_i\in \mathbb{R}^3$ \why{represented as view-dependent radiance encoded via spherical harmonics (SH) coefficients}.

To render an image, 3DGS \why{first projects each Gaussian $G_i$ onto the 2D image plane through} perspective transformation, \why{resulting in 2D Gaussians}.
\ryn{It then} sorts $\mathcal{N}$ Gaussians \why{that overlap the target pixel} by depth, and \ryn{computes} the \why{final pixel color $C$ using $\alpha$-blending}:
\vspace{-2mm}
\begin{equation}
    C = \sum_{i\in \mathcal{N}} c_i \eta_i \prod_{j=1}^{i-1} (1-\eta_j),
\end{equation}
where $\eta$ is given by \why{evaluating the
2D Gaussian with covariance matrix multiplied with a
learned per-point opacity.}

\subsection{\bracketgs}
We draw inspiration from Endo~\etal's single-image HDR reconstruction method~\cite{endo_2017_tog_drtmo}, which synthesizes multi-exposure variants from a single LDR input for HDR reconstruction. 
We explore this idea in 3D by proposing \bracketgs,
which estimates multi-exposure \why{3D Gaussians} and reconstructs an HDR \why{3D Gaussians from them}.
Unlike standard 3DGS, our \why{Bracketed and HDR 3D Gaussians} $\mathcal{G}$ integrates explicit exposure modeling, which can be formulated as:
\vspace{-1mm}
\begin{align}
    \begin{split}
        \mathcal{G} =& \{G_i (\mu_i, \Sigma_i, \alpha_i, \{c_i^{l_1}, \dots, c_i^{l_n}, c_i^h\}, \\
        &\{\Delta t_i^1, \dots, \Delta t_i^n\}, \theta, \sigma) |\ i=1, 2, \dots, N\},
    \end{split}
\end{align}
where $c_i^h$ denotes the HDR radiance, while $\{c_i^{l_1}, \dots, c_i^{l_n}\}$ correspond to linear radiance values \ryn{of} exposures $\{\Delta t_i^1, \dots, \Delta t_i^n\}$.
$\theta$ represents the MLP parameters, and $\sigma$ represents the standard deviations of well-exposedness, both functioning in \ryn{the} Differentiable Neural Exposure Fusion module and are shared across all Gaussians.
Note that we parameterize linear radiance for 3D Gaussians instead of LDR color values~\cite{Cai_2024_nips_hdrgs} to facilitate exposure manipulation capabilities.

Following the imaging model in \ryn{Eq.~\ref{eq:image-formation}}, we assume that scene radiance $E$ is constant for all exposures and treat $E\Delta t$ as a linear radiance space proportional to exposure time, yielding:
\vspace{-2mm}
\begin{align}
    c_i^h = \frac{c_i^{l_1}}{\Delta t_i^1} = \dots = \frac{c_i^{l_n}}{\Delta t_i^n}.
\end{align}
Given a base Gaussian parameterized by linear radiance $c_i^{l_1}$ at input exposure $\Delta t_i^1$, we sample additional exposures $\Delta t_i^2, \dots, \Delta t_i^n$. The corresponding linear radiance values are derived via:
\vspace{-2mm}
\begin{align}
    c_i^{l_j} = c_i^{l_1}\frac{\Delta t_i^j}{\Delta t_i^1}, \quad 2 \leq j \leq n.
\end{align}

Based on the tradition of non-parametric CRF calibration principles~\cite{debevec_1997_siggraph_hdr, Cai_2024_nips_hdrgs}, we further enhance training stability by transforming linear radiance values into logarithmic space:
\vspace{-3mm}
\begin{align}
    \log{c_i^{l_j}} = \log{c_i^{l_1}} + \log{\Delta t_i^j} - \log{\Delta t_i^1},
\end{align}
where $\log(\cdot)$ denotes the natural logarithm.
The linear radiance $c_i^{l_j}$ can be easily recovered through exponentiation $c_i^{l_j} = \exp(\log{c_i^{l_j}})$.
Ultimately, bracketed Gaussians is obtained by aggregating the derived linear radiance values $c_i^{l_j}$ across varying exposure times and $c_i^{l_1}$ in base Gaussian, thereby constructing a comprehensive multi-exposure representation of the 3D Gaussian scene.

\subsection{Differentiable Neural Exposure Fusion \ryn{(NeEF)}}

\begin{figure}[th]
    \centering
    \includegraphics[width=\linewidth]{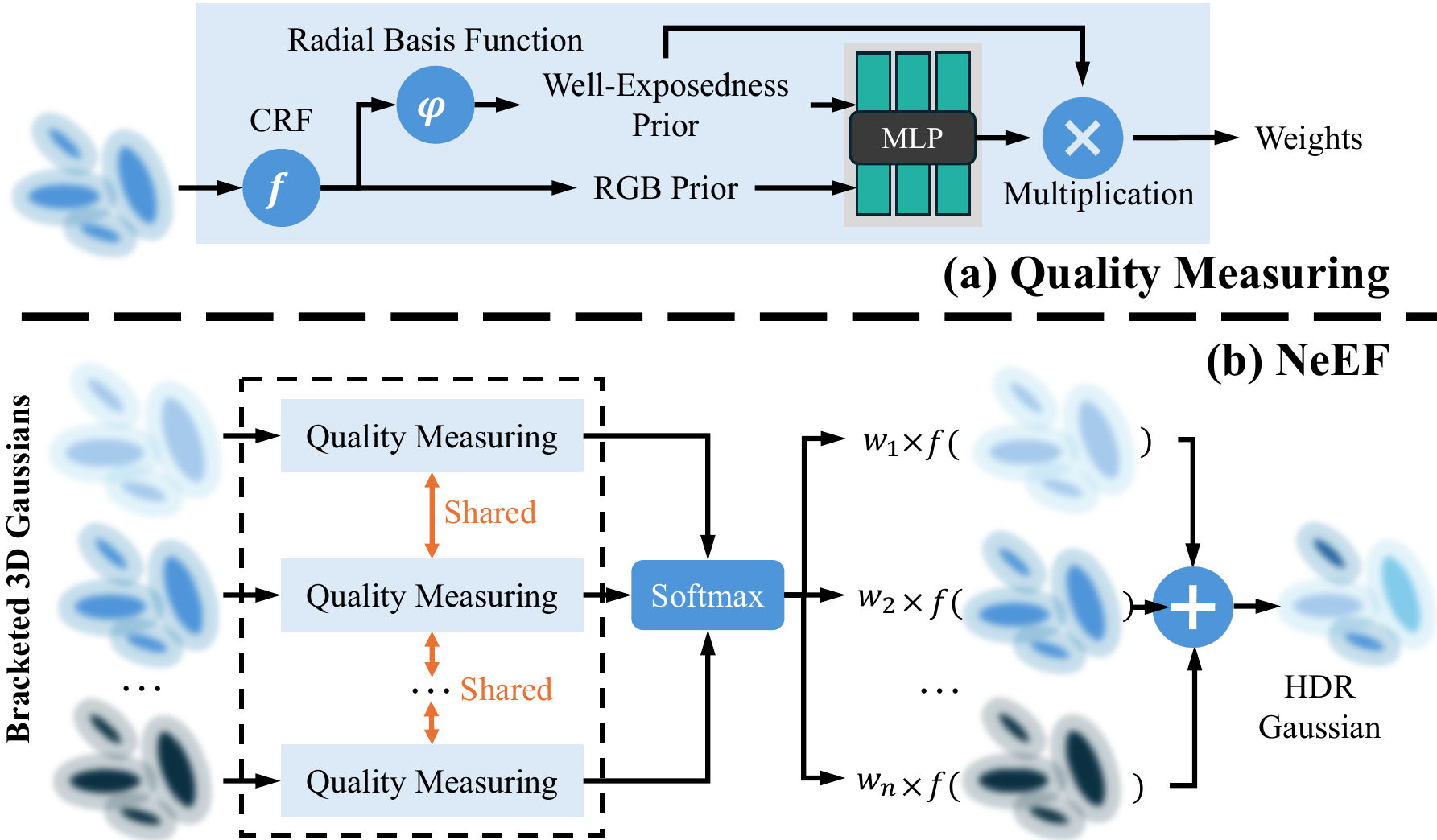}
    \caption{The structure of Differentiable Neural Exposure Fusion Module.}
    \vspace{-3mm}
    \label{fig:neef}
\end{figure}

We now aim to fuse bracketed Gaussians into a unified HDR Gaussian representation.
We note that conventional exposure fusion techniques \why{primarily operate in 2D space~\cite{mertens_2007_cga_exposure_fusion} and heavily rely on} spatial information. These methods are inherently incompatible with 3D Gaussian representations, \why{as Gaussian point clouds, unlike structured 2D-pixel grids, lack inherent spatial coherence}.

To address \why{this challenge}, we propose Neural Exposure Fusion (NeEF), a differentiable module that adaptively fuses multi-exposure Gaussians into an \why{HDR Gaussians} through learnable parameters.
While adhering to the weighted fusion paradigm of traditional exposure fusion methods~\cite{mertens_2007_cga_exposure_fusion}, we fundamentally reformulate the weight computation mechanism: instead of deriving weights from 2D local features (\eg, contrast or gradients), we leverage neural predictions from a MLP, enabling adaptive fusion.
% , inspired by recent neural exposure fusion works~\cite{Onzon_2021_CVPR_neural_exposure_fusion, Onzon_2024_CVPR_neural_exposure_fusion}.

Specifically, to align with conventional exposure fusion workflows that operate in \ryn{the} LDR color space, we first apply a fixed tone-mapping function $f(\cdot)$ to convert linear radiance values $c_i^l$ into LDR colors $f(c_i^l)$ for subsequent processing.
The tone mapper $f(\cdot)$ is selected in advance from the Database of Response Functions (DoRF)~\cite{grossberg_2004_pami_emor} for each scene individually. 
%use a CRF estimation method
We select the best-performing CRF out of DoRF based on their performance on a randomly sampled training image, evaluated by statistical metrics, \ie, dynamic range and smoothness.
%~\kk{(**This sentence is too long. Try break it.**)}
Once selected, $f(\cdot)$ remains fixed throughout the whole training process to ensure numerical stability.

\vspace{1ex}

\noindent{\bf Exposure Quality Measuring.}
As shown in \cref{fig:neef}(a), we measure the exposure quality of $i$-th 3D Gaussian with $p$-th exposure using a per-channel well-exposedness metric $\gamma_{i,p}^c$, where $c\in \{R, G, B\}$ and $1\leq p\leq n$ enumerates $n$ exposure times.
This metric quantifies the exposure quality of each Gaussian, prioritizing those with radiance values \yy{that are neither under-exposed ($\approx$ 0) nor over-exposed ($\approx$ 1).}
%\rynq{that avoid (*** Do you mean "to avoid"? ***)} \ryn{under-exposure} ($\approx$ 0) or \ryn{over-exposure} ($\approx$ 1).
%
To implement this, we compute weights via a radial basis function $\varphi(\cdot)$ applied to each color channel:
\begin{align}
    \gamma_{i,p}^c = \varphi(f(c_i^{l_p, c})) = \exp\left( -\frac{(f(c_i^{l_p, c}) - 0.5)^2}{2\sigma_c^2} \right),
\end{align}
where $c_i^{l_p, c}$ denotes the linear radiance of the $c$-channel (R, G, B) for the $i$-th Gaussian with $p$-th exposure.
Crucially, we also avoid excessively penalizing under- or over-exposed Gaussians, as these extremes often retain critical scene details: shadows in over-exposed Gaussians and highlights in under-exposed ones.
To balance detail preservation with exposure quality, we parameterize the standard deviations $\{\sigma_R, \sigma_G, \sigma_B\}$ as learnable weights initialized to 0.2.
This design offers the model optimization-driven flexibility \why{to adaptively tune exposure sensitivity per channel adaptively}, ensuring robust highlight recovery and shadow detail retention during HDR reconstruction.

\vspace{1ex}

\noindent{\bf Weighted Fusion.}
To account for channel-dependent variations in the camera response function (CRF), we employ three independent MLPs, denoted collectively as $\theta$, to compute per-channel fusion weights.
These MLPs adaptively prioritize Gaussians based on their exposure characteristics in each RGB channel.
For the $i$-th Gaussian, $p$-th exposure, the final weight $w_{i}^c\in \mathbb{R}^{n\times 1}$ in channel $c$ is computed as:
\vspace{-4mm}
\begin{align}
    w_{i, p}^c =  \theta_c(\gamma_i^c, f(c_i^{l_p, c}))\otimes \gamma_i^c ,\\
    w_{i}^c = \text{softmax} \left( \left[ w_{i, p}^c \right]_{p=1}^{n} \right)
\end{align}
where $\otimes$ denotes point-wise multiplication, and $\theta_c$ is the MLP for channel $c$, which outputs $w_{i, p}^c$.
\why{We then stack it across multi-exposures to get global weights and perform softmax operation along the exposure dimension to produce $w_{i}^c$}.
In this progress, $\theta_c$ processes the well-exposedness metric $\gamma_i^c$ and the LDR-mapped value $f(c_i^{l, c})$ to refine the fusion weight, ensuring channel-specific adaptation to CRF disparities.

% \why{(Softmax should be for fusing across p dimension instead of points num dimension.)}

The final HDR Gaussian radiance $c_i^h$ is composited through a weighted summation of tone-mapped LDR Gaussians $f(c_i^{l})$ within the NeEF module, as shown in \cref{fig:neef}(b):
\vspace{-4mm}
\begin{align}
    c_i^h = \sum_{p=1}^{n} \left[ w_{i, p'}^c f(c_i^{l_p, c}) \right]_{c\in \{R, G,B\}},
\end{align}
where we first calculate \why{the weighted sum along the exposure axis} and then concatenate RGB channels back to produce HDR \why{color} $c_i^h$.
This integration preserves critical scene details across exposure extremes while ensuring photometric consistency for faithful HDR reconstruction.

\vspace{1ex}

\noindent{\bf Rendering.}
In contrast to HDR-GS~\cite{Cai_2024_nips_hdrgs}, which models the logarithmic domain of HDR radiance, our framework directly models HDR radiance $c_i^h$ in linear space using spherical harmonics (SH) coefficients:
\begin{align}
    C_\text{SH} = \{k_l^m|0\leq l\leq L, -l\leq m\leq l\}\in \mathbb{R}^{(L+1)^2 \times 3},
\end{align}
where $L$ denotes the maximum SH degree.
This design choice is necessitated by the output characteristics of NeEF, which operates natively in linear radiance space, thereby eliminating the need for domain conversion.
Each $k_l^m \in \mathbb{R}^3$ corresponds to RGB-specific coefficients for the SH basis of degree $l$ and order $m$.
During rendering, the view-dependent HDR radiance $c_i^h$ at direction $\mathbf{d}=(\theta, \phi)$ is: % computed as:
\vspace{-2mm}
\begin{align}
    c_i^h(\mathbf{d}, C_\text{SH}) = \sum_{l=0}^L \sum_{m=-l}^{l} k_l^m Y_l^m(\theta, \phi),
\end{align}
where $Y_l^m(\theta, \phi):\mathbb{S}^2\rightarrow \mathbb{R}$ represents the SH basis function that maps spherical coordinates on the unit sphere $\mathbb{S}^2$ to real values.

\subsection{Optimization}

\noindent \textbf{LDR Reconstruction Loss.}
The training objective for \papertitle is formulated as a weighted combination of the mean absolute error ($\mathcal{L}_1$) and structural dissimilarity ($\mathcal{L}_\text{D-SSIM}$) between rendered and ground-truth LDR views.
Given camera intrinsic matrices $\mathbf{K}\in \mathbb{R}^{3\times 3}$ and extrinsic matrices $\mathbf{E} = [\mathbf{R} | t]\in \mathbb{R}^{3\times 4}$, we splat 3D Gaussians onto the 2D image plane, sort them by depth, and perform rasterization to synthesize LDR views as:
\begin{align}
    % \mathbf{I}^h = F \left(\mathbf{K}, \mathbf{E}, \{\mu_i, \Sigma_i, \alpha_i, c_i^h\}_{i=1}^{N} \right),\\
    \mathbf{I}^l = F \left(\mathbf{K}, \mathbf{E}, \{\mu_i, \Sigma_i, \alpha_i, f(c_i^h\Delta t_1)\}_{i=1}^{N} \right),
\end{align}
where $F$ denotes the rasterization function.
The LDR reconstruction loss is computed as:
\begin{align}
    \ell_\text{LDR} = \mathcal{L}_1 (\mathbf{I}^l, \hat{\mathbf{I}}^l) + \lambda \mathcal{L}_\text{D-SSIM} (\mathbf{I}^l, \hat{\mathbf{I}}^l),
\end{align}
where $\lambda$ is a weighting hyperparameter \why{that balances the two loss terms}, and $\hat{\mathbf{I}}^l$ represents the ground-truth LDR image.

\vspace{1ex}

\noindent \textbf{Linear Loss.}
To ensure consistency in the linear radiance space $c_i^l$, we introduce an auxiliary linear loss $\ell_\text{Linear}$ \why{that supervises intermediate reconstructions}.
For synthetic scenes with available HDR ground truth $\hat{\mathbf{I}}^h$, we directly supervise $c_i^l$ by aligning the rendered view of base Gaussian with scaled HDR radiance values $\hat{\mathbf{I}}^h \times \Delta t_1$.
However, in real-world scenarios \why{where HDR ground truth is unavailable}, the \why{unsupervised optimization of $c_i^l$ frequently leads to corrupted radiance estimations and failed HDR reconstructions}.
To mitigate this, we employ a self-supervised strategy where novel views rendered from the fused HDR Gaussians \why{guide the linear radiance outputs of the base Gaussians. Specifically:}
\begin{align}
    \mathbf{I}^h = F \left(\mathbf{K}, \mathbf{E}, \{\mu_i, \Sigma_i, \alpha_i, c_i^h\}_{i=1}^{N} \right),\\
    \mathbf{I}^{b} = F \left(\mathbf{K}, \mathbf{E}, \{\mu_i, \Sigma_i, \alpha_i, c_i^{l_1}\}_{i=1}^{N} \right),
\end{align}
where 
% $F$ denotes the rasterization function, 
$\mathbf{I}^h$ represents novel views rendered from the fused HDR Gaussians, and $\mathbf{I}^{b}$ corresponds to views rendered from the base Gaussians in linear radiance space,
the linear loss is computed as:
\begin{align}
    \ell_\text{Linear} = \mathcal{L}_1 (\mathbf{I}^h \Delta t_1, \mathbf{I}^b) + \lambda \mathcal{L}_\text{D-SSIM} (\mathbf{I}^h \Delta t_1, \mathbf{I}^b).
\end{align}

The whole training objective can be formulated as:
\begin{align}
    \ell = \ell_\text{LDR} + \beta \ell_\text{Linear},
\end{align}
where $\beta=0.5$, \why{which balances} the contribution of LDR reconstruction fidelity and linear radiance consistency.

\section{Experiments}
\label{sec:exp}

\subsection{Experimental Settings}

\noindent \textbf{Datasets.}
Following established \why{benchmarks for HDR scene reconstruction}~\cite{huang_2022_cvpr_hdrnerf, Cai_2024_nips_hdrgs}, we evaluate our method on the HDR-NeRF dataset~\cite{huang_2022_cvpr_hdrnerf}, comprising four real-world scenes captured with camera and eight synthetic scenes rendered in Blender~\cite{blender}.
Each scene includes 35 viewpoints, with five exposures $\{t_1, t_2, t_3, t_4, t_5\}$ per viewpoint.
In contrast to multi-exposure methods~\cite{huang_2022_cvpr_hdrnerf, Cai_2024_nips_hdrgs}, which utilize multiple exposures during training, \why{we adopt a single-exposure training strategy. Specifically, we conduct five separate training runs per scene, one for each exposure ($t_1$ through $t_5$), and report averaged metrics across all exposure settings to ensure robust performance evaluation.} 
% we train our model separately on each single exposure ($t_1$ to $t_5$), performing five independent training runs per scene.
%
% Reported metrics are averaged across all exposure configurations to ensure robustness.

For evaluation, we \why{follow the HDR-NeRF dataset’s standard split}: 18 views for training and 17 held-out views for testing.
Quantitative \why{experiments are conducted} on the synthetic subset, where HDR ground truth is available for metric calculation.
Qualitative comparisons are performed on the real-world HDR-NeRF scenes.
% and MipNeRF 360 dataset~\cite{barron_2022_cvpr_mipnerf360}

\noindent \textbf{Metrics.}
We quantitatively evaluate reconstructed HDR novel views using standard metrics: Peak Signal-to-Noise Ratio (PSNR), Structural Similarity Index (SSIM)~\cite{wang_2004_TIP_SSIM}, and Learned Perceptual Image Patch Similarity (LPIPS)~\cite{zhang_2018_cvpr_lpips}.
Since HDR imagery requires tone-mapping for display, we follow the evaluation protocol of HDR-NeRF~\cite{huang_2022_cvpr_hdrnerf} and HDR-GS~\cite{Cai_2024_nips_hdrgs}, computing metrics in the tone-mapped domain via the $\mu$-law operator, a canonical compression function widely adopted in HDR benchmarking~\cite{Kalantari_2017_tog_hdr, Prabhaker_2020_eccv_hdr, yan_2019_cvpr_attention_hdr}:
\vspace{-4mm}
\begin{align}
    M(E) = \frac{\log(1+\mu E)}{\log(1+\mu)},
\end{align}
where $E$ denotes normalized HDR radiance values, and $\mu = 5000$ controls the dynamic range compression rate.

\subsection{Quantitative Results}

\begin{figure*}[ht]
\renewcommand{\tabcolsep}{0.8pt}
\renewcommand{\arraystretch}{0.7}
    
    \begin{center}
        \resizebox{\linewidth}{!}{
        \begin{tabular}{ccccccc}
            \includegraphics[width=0.14\linewidth]{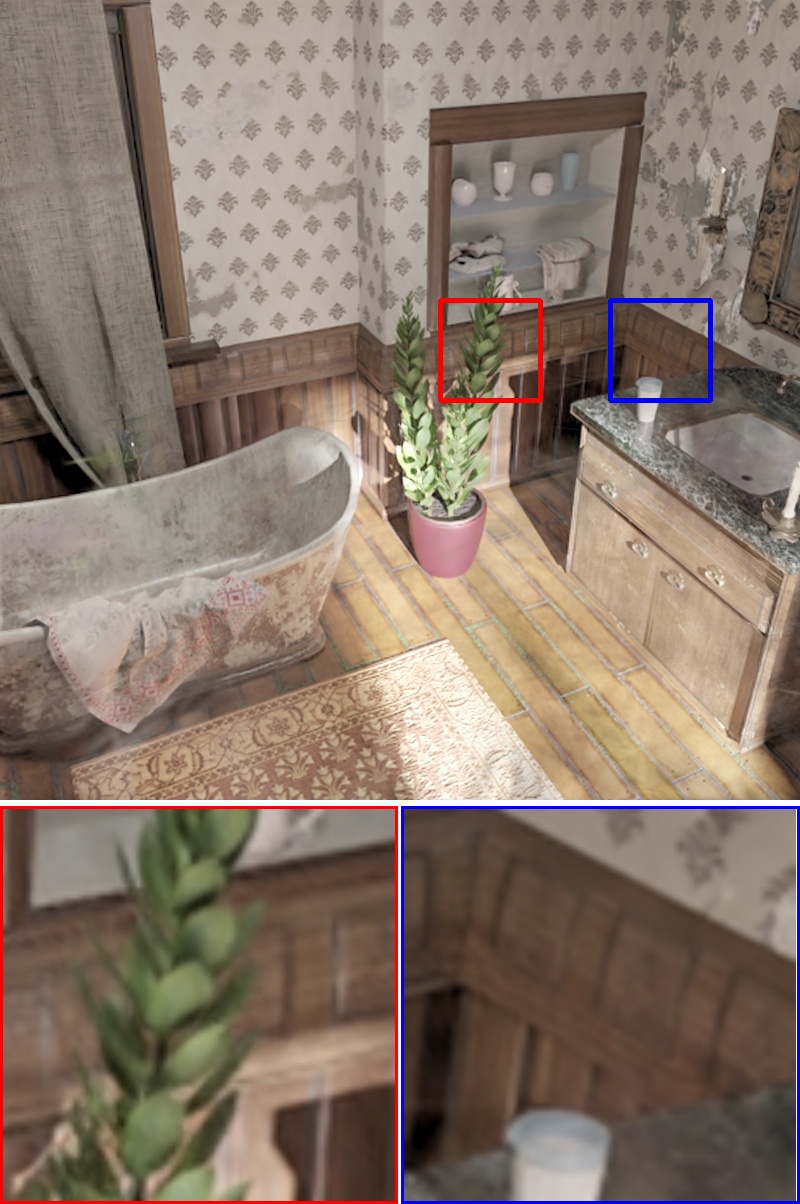} &
            \includegraphics[width=0.14\linewidth]{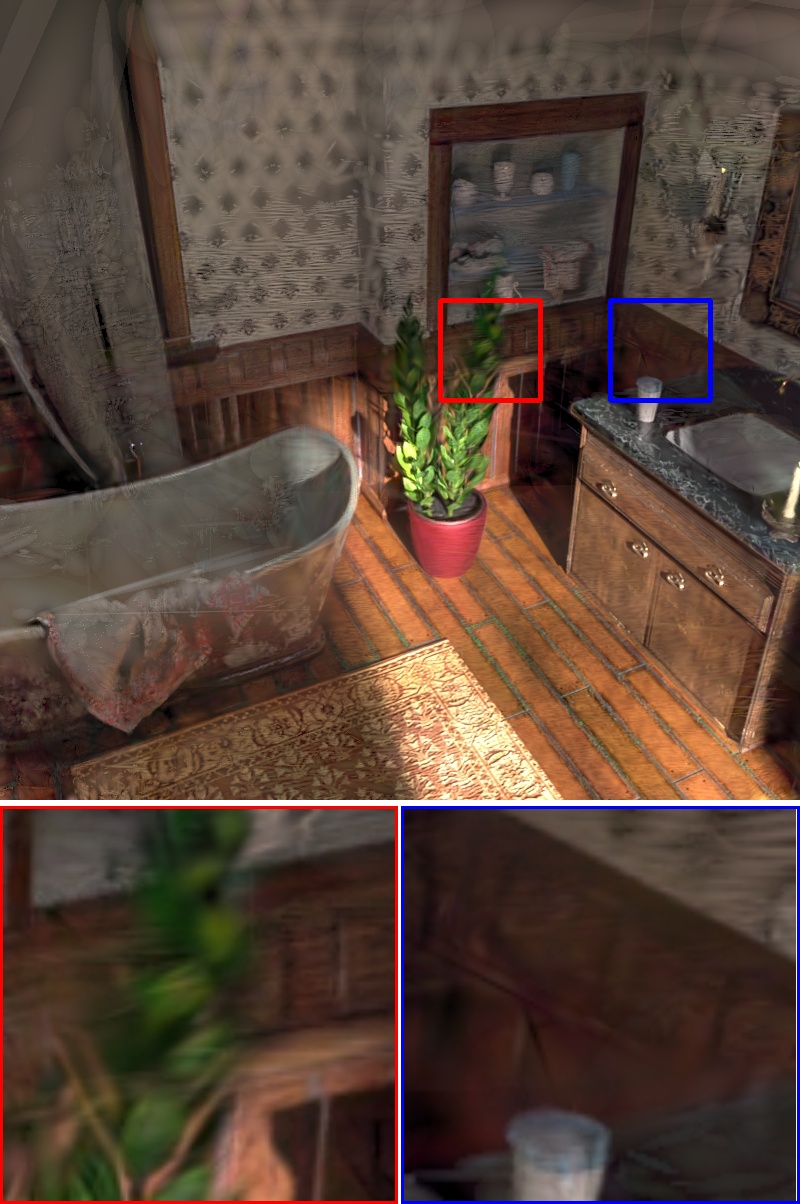} &
            \includegraphics[width=0.14\linewidth]{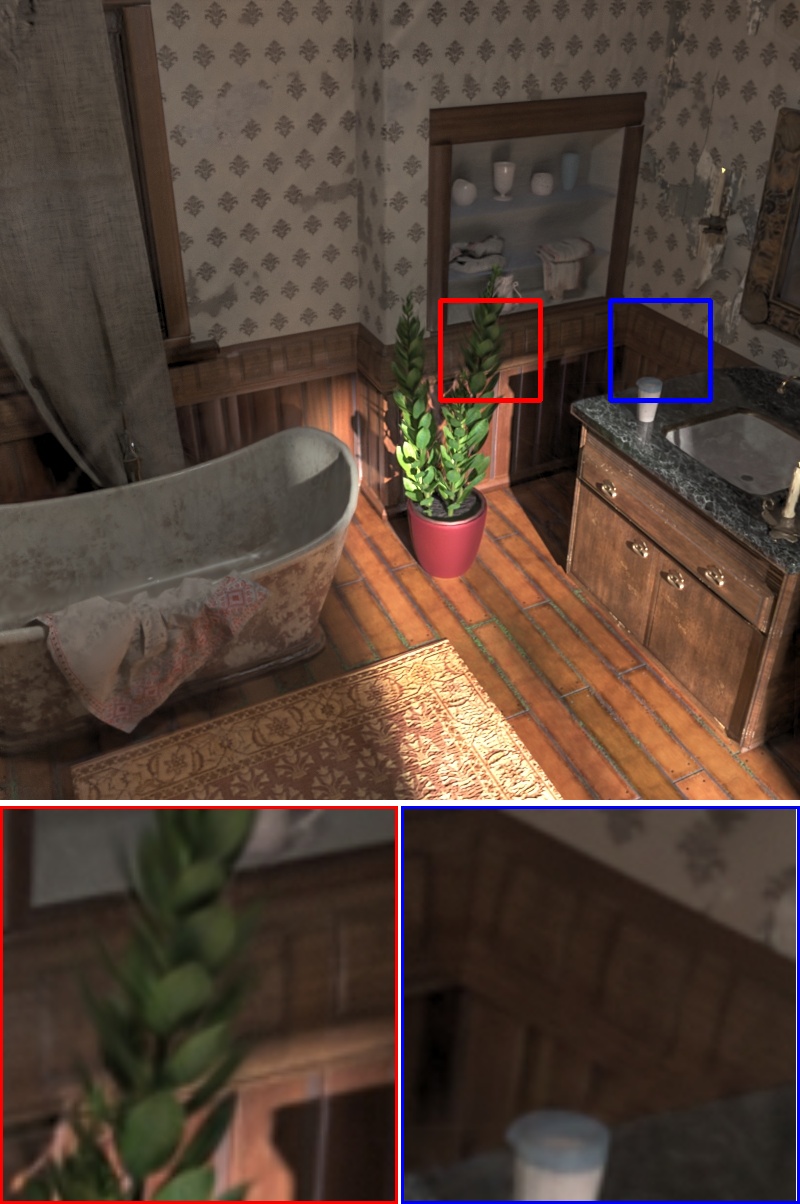} &
            \includegraphics[width=0.14\linewidth]{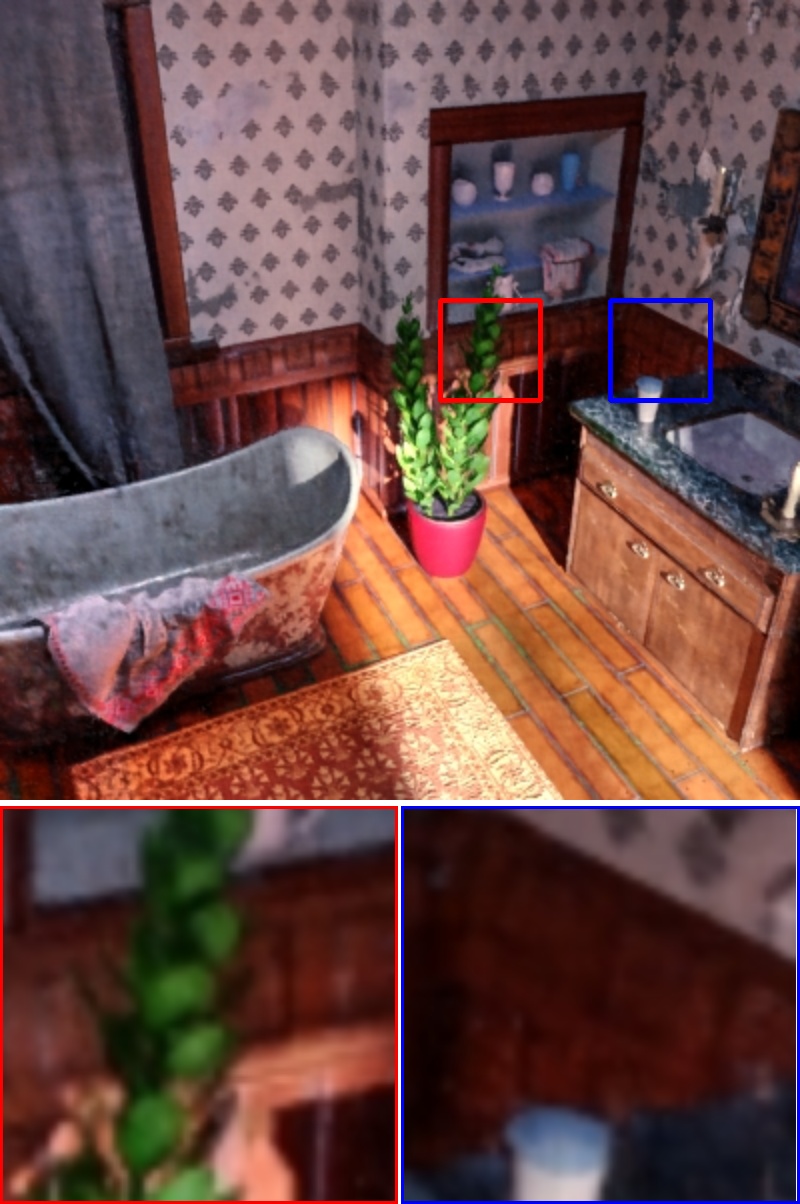} &
            \includegraphics[width=0.14\linewidth]{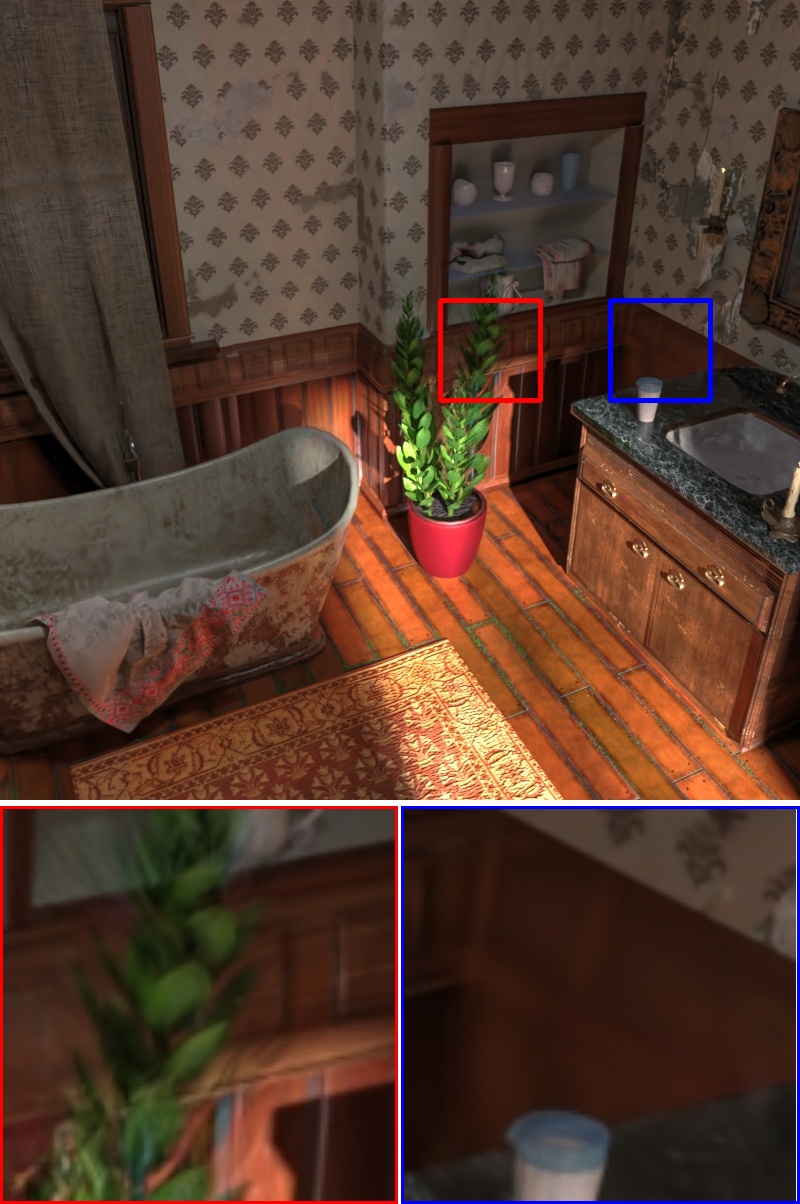} &
            \includegraphics[width=0.14\linewidth]{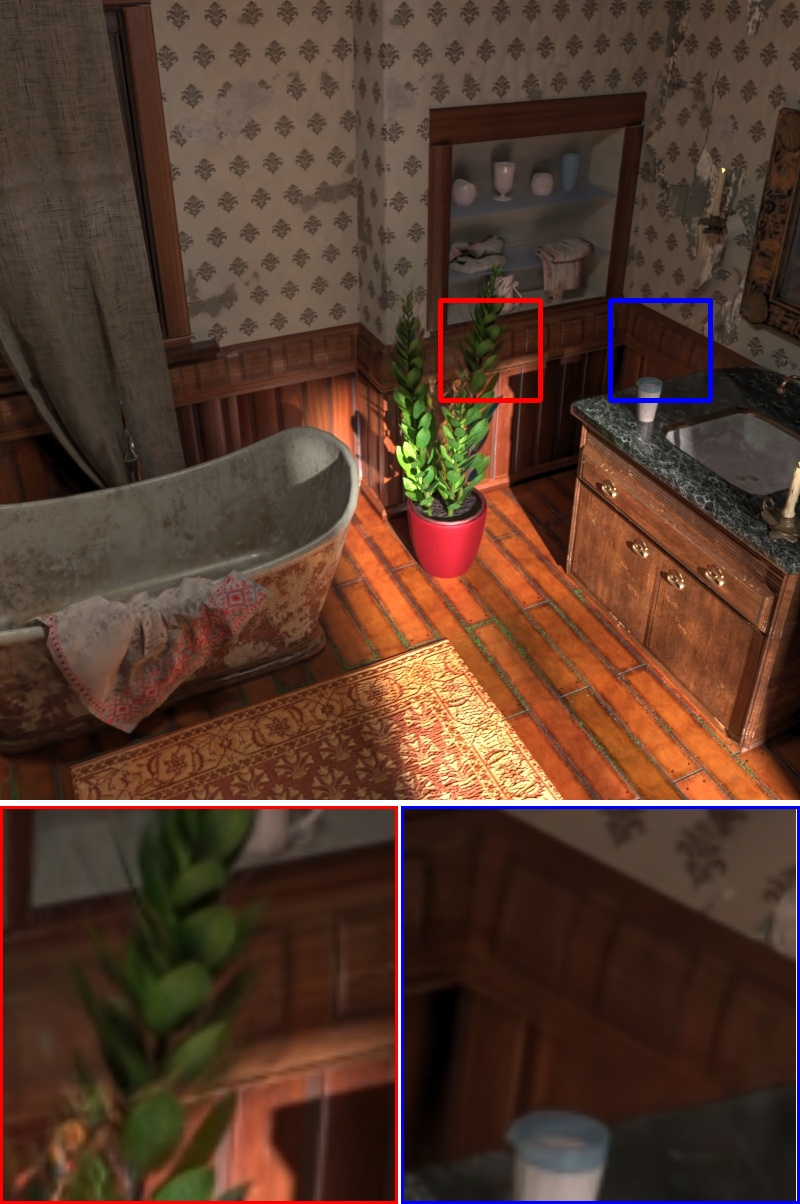} &
            \includegraphics[width=0.14\linewidth]{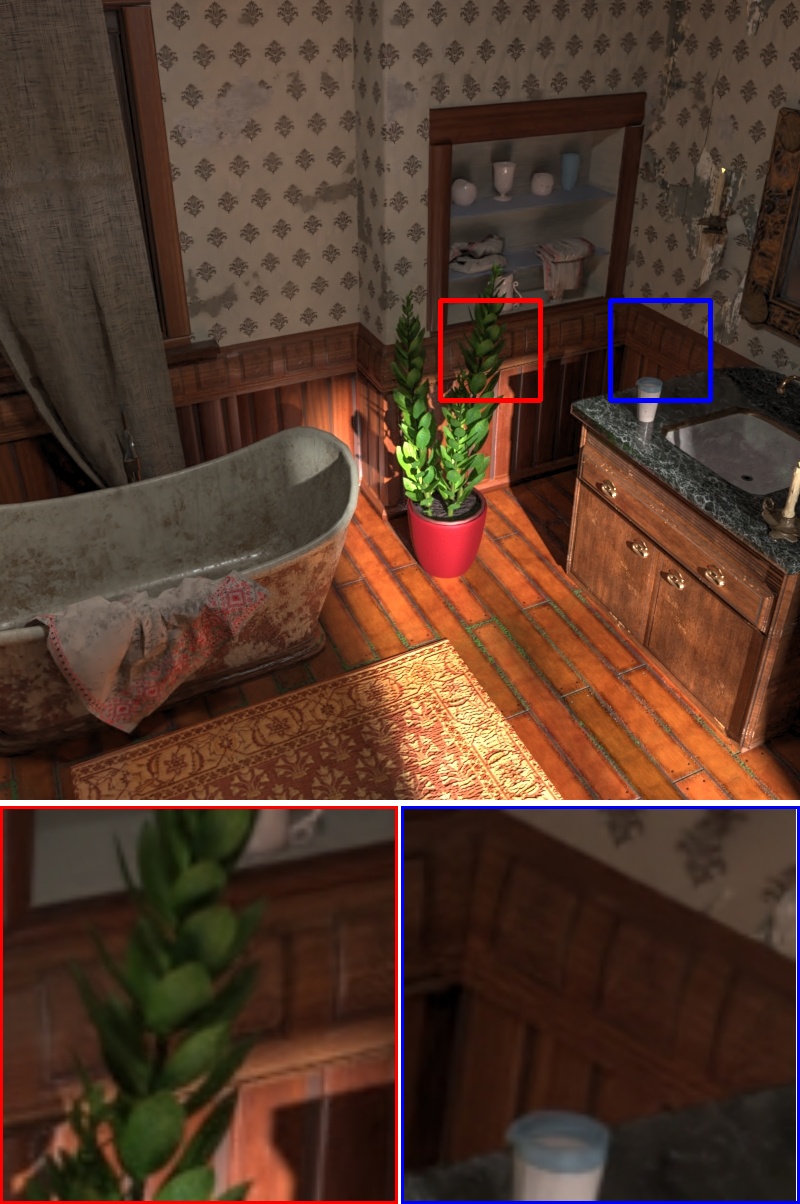} \\

            \scriptsize{DrTMO~\cite{endo_2017_tog_drtmo}+GS} & \scriptsize{SingleHDR~\cite{Liu_2020_CVPR_singlehdr}+GS} &
            \scriptsize{IntrinHDR~\cite{dille_2024_eccv_IntrinsicHDR}+GS} &
            \scriptsize{HDR-NeRF~\cite{huang_2022_cvpr_hdrnerf}} &
            \scriptsize{HDR-GS~\cite{Cai_2024_nips_hdrgs}} &
            \scriptsize{Ours} &
            \scriptsize{Ground Truth} \\

            \includegraphics[width=0.14\linewidth]{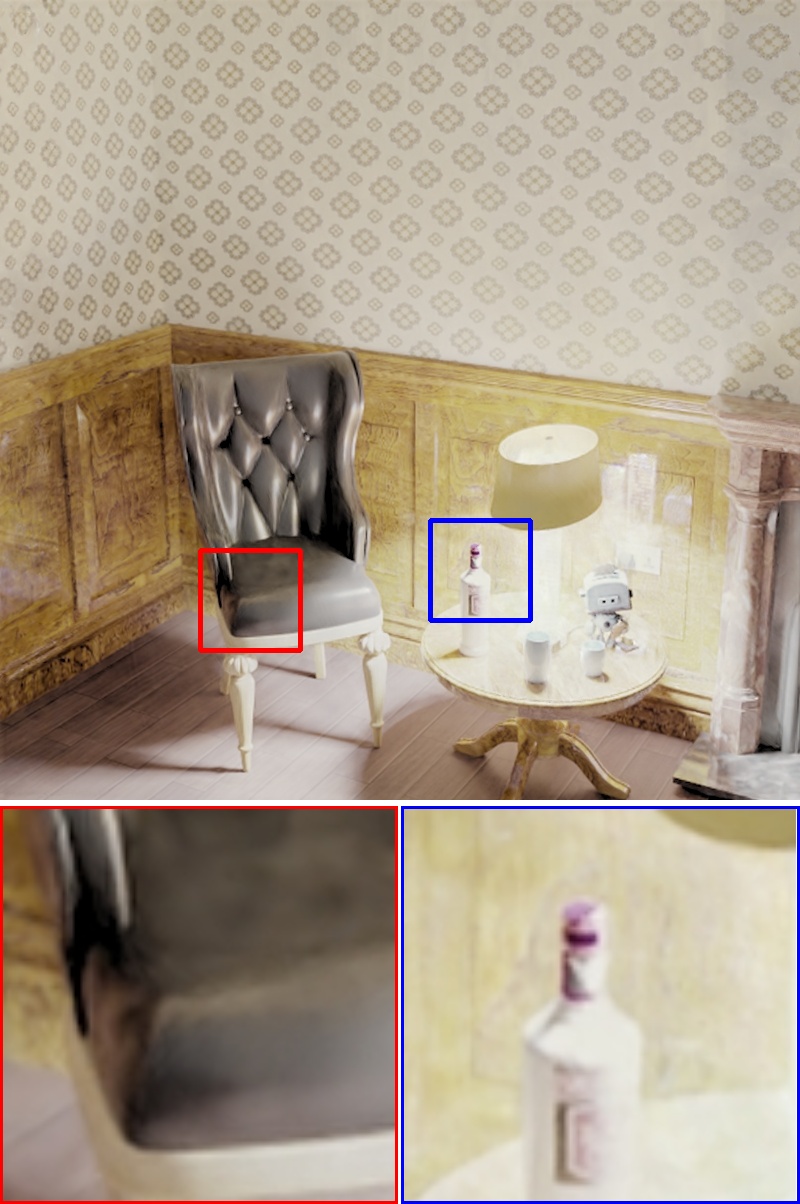} &
            \includegraphics[width=0.14\linewidth]{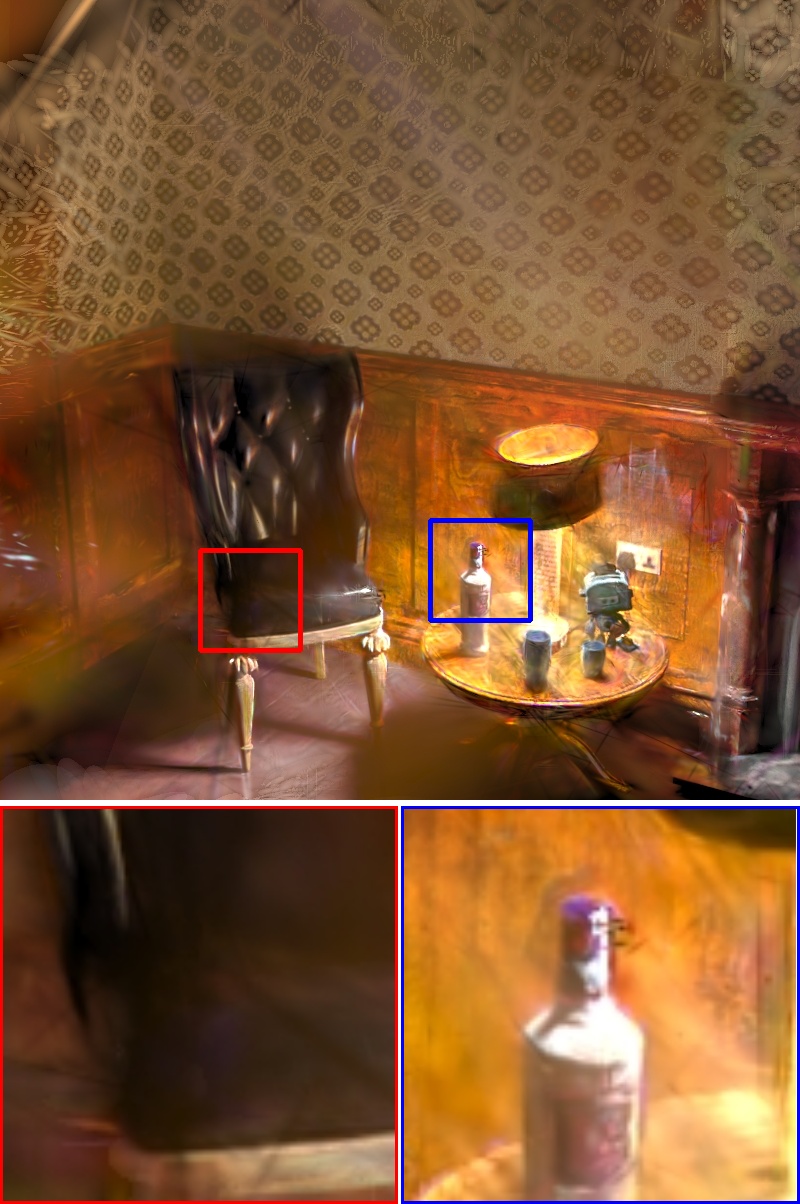} &
            \includegraphics[width=0.14\linewidth]{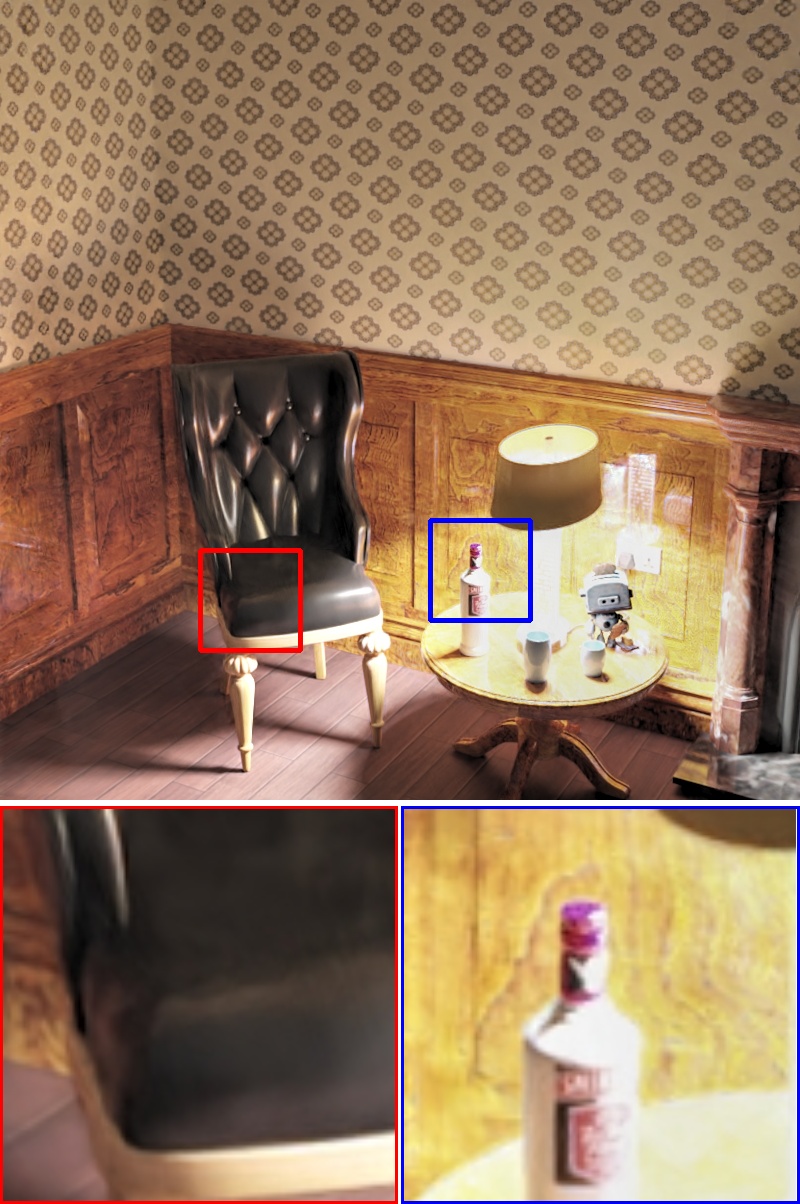} &
            \includegraphics[width=0.14\linewidth]{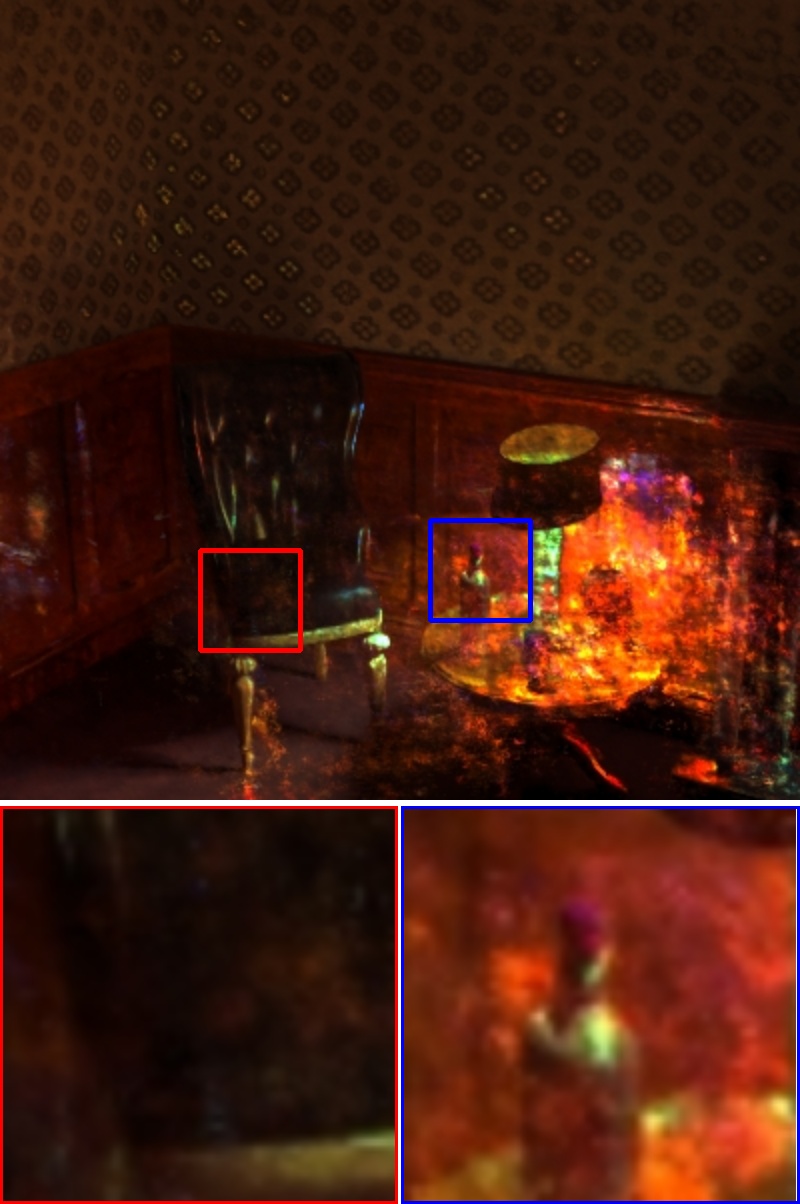} &
            \includegraphics[width=0.14\linewidth]{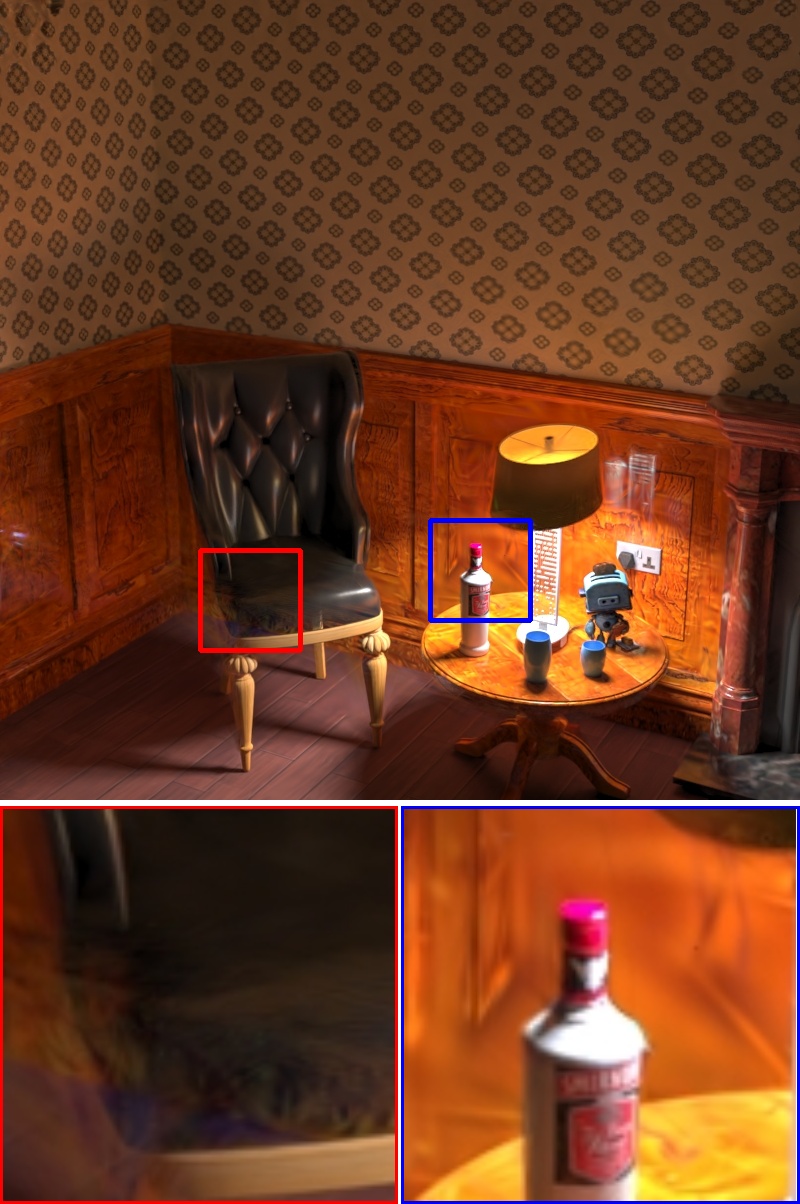} &
            \includegraphics[width=0.14\linewidth]{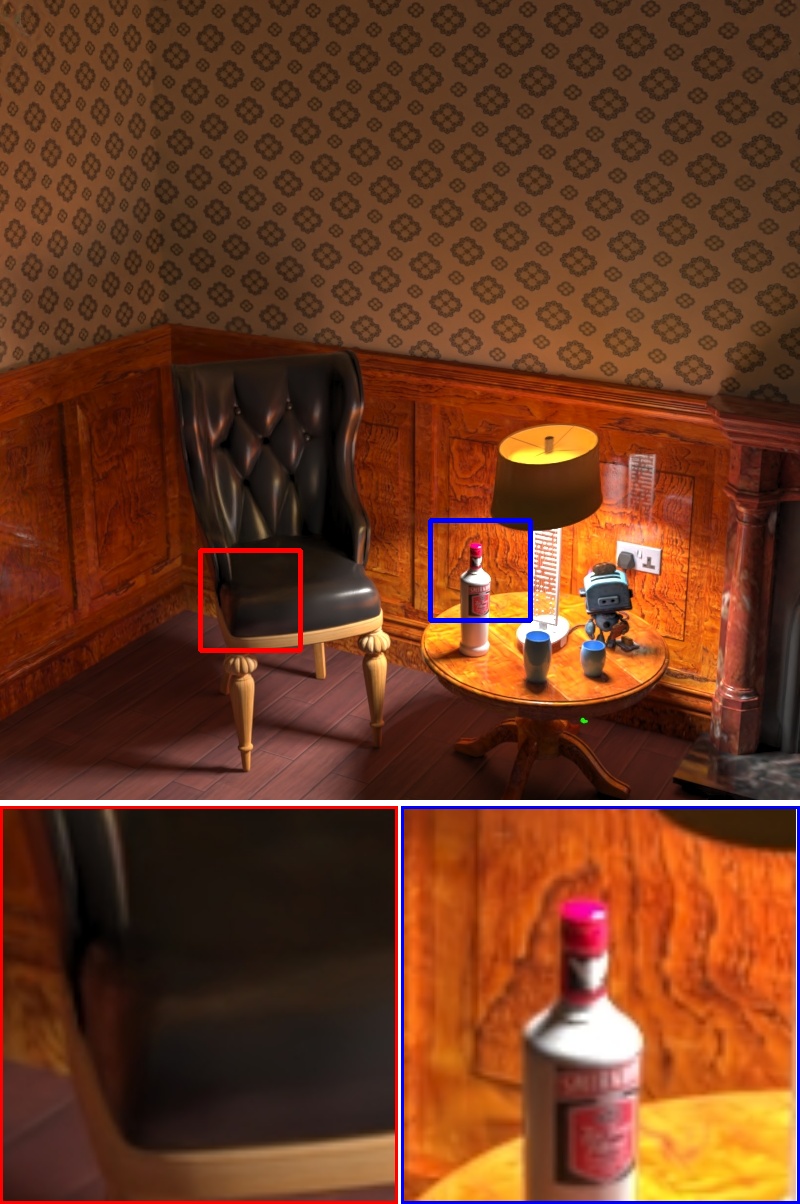} &
            \includegraphics[width=0.14\linewidth]{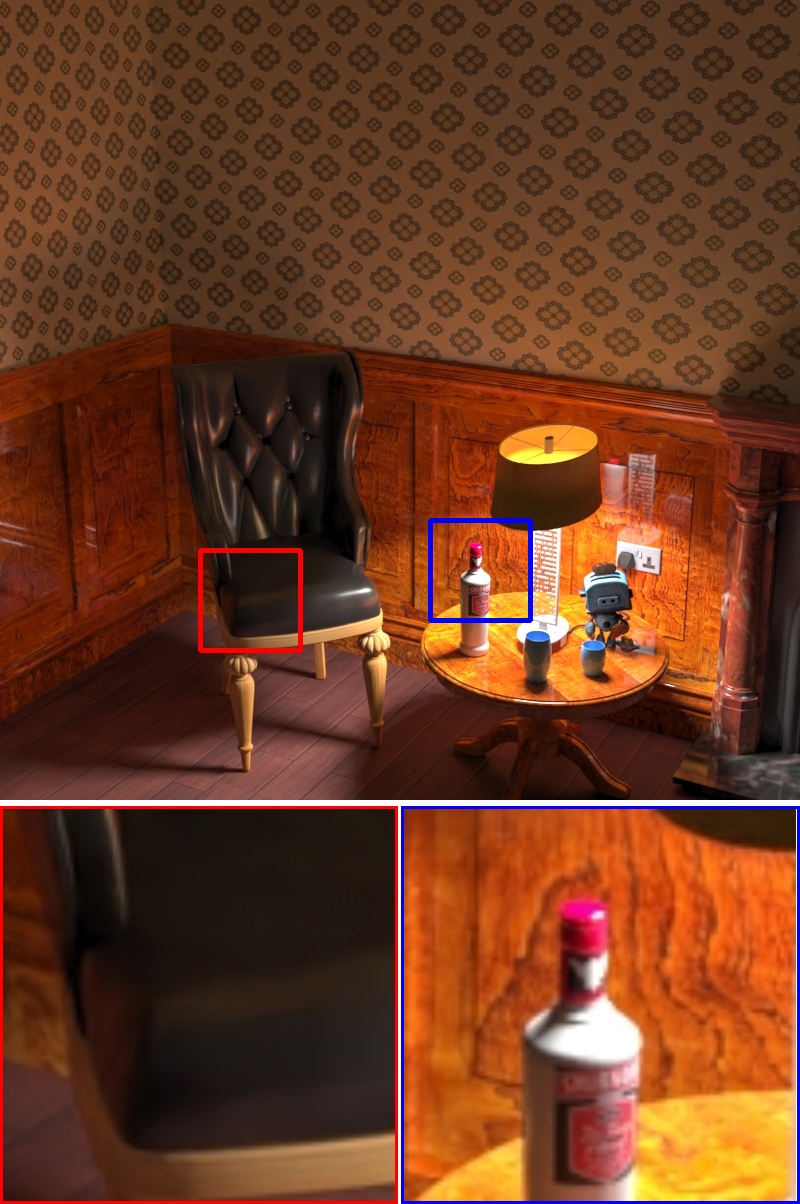} \\

            \scriptsize{DrTMO~\cite{endo_2017_tog_drtmo}+GS} & \scriptsize{SingleHDR~\cite{Liu_2020_CVPR_singlehdr}+GS} &
            \scriptsize{IntrinHDR~\cite{dille_2024_eccv_IntrinsicHDR}+GS} &
            \scriptsize{HDR-NeRF~\cite{huang_2022_cvpr_hdrnerf}} &
            \scriptsize{HDR-GS~\cite{Cai_2024_nips_hdrgs}} &
            \scriptsize{Ours} &
            \scriptsize{Ground Truth}
            
        \end{tabular}
        }
    \vspace{-3mm}
    \caption{Qualitative comparison on synthetic scenes with HDR ground truth. Combining image HDR reconstruction methods with 3DGS tends to produce novel views with floaters, while multi-exposure HDR-NVS methods on single-exposure scenarios tend to produce unrealistic colors due to lack of multi-exposure or blurry regions due to misalignment of single exposures. All the HDR results are tone mapped using PhotoMatix~\cite{photomatixpro} under the same settings for visualization.}

    \vspace{-4mm}
    \label{fig:visual}
    \end{center}
    
\end{figure*}

\begin{table}[tb]
\renewcommand{\arraystretch}{1.4}
    \centering
    \resizebox{.9\linewidth}{!}{
    \begin{tabular}{l|ccc}
    \noalign{\hrule height2pt}
        Methods & PSNR~$\uparrow$ & SSIM~$\uparrow$ & LPIPS~$\downarrow$ \\
        \hline
        DrTMO~\cite{endo_2017_tog_drtmo}+GS & 4.640 & 0.407 & 0.610 \\
        SingleHDR~\cite{Liu_2020_CVPR_singlehdr}+GS & 14.251 & 0.629 & 0.347 \\
        IntrinHDR~\cite{dille_2024_eccv_IntrinsicHDR}+GS & 8.201 & 0.556 & 0.367 \\
        \hline
        HDR-NeRF~\cite{huang_2022_cvpr_hdrnerf} & 12.581 & 0.366 & 0.608 \\
        HDR-GS~\cite{Cai_2024_nips_hdrgs}~(w/~GT) & 36.821 & 0.965 & 0.046 \\
        HDR-GS~\cite{Cai_2024_nips_hdrgs}~(w/o~GT) & 11.008 & 0.516 & 0.442 \\
        \hline
        \papertitle(w/~GT) & 37.054 & 0.965 & 0.032 \\
        \papertitle(w/o~GT) & 25.371 & 0.810 & 0.117 \\
    \noalign{\hrule height2pt}
    \end{tabular}}
    \vspace{-2mm}
    \caption{Quantitative results for rendered HDR novel views trained on HDR-NeRF synthetic dataset under single exposure settings are reported. HDR novel views are evaluated against the HDR ground truth. The reported values represent averages across all eight synthetic scenes and five exposure levels. The notation ``w/~GT'' and ``w/o~GT'' indicate training with or without HDR ground truth, respectively.}
    \vspace{-5mm}
    \label{tab:quantitative-avg}
\end{table}

We report the quantitative results of HDR novel view synthesis in \cref{tab:quantitative-avg}.
Our proposed method \papertitle is compared against two categories of baselines:
\begin{enumerate*}[label={(\arabic*)}]
    \item multi-exposure-based HDR-NVS methods, specifically HDR-NeRF~\cite{huang_2022_cvpr_hdrnerf} and HDR-GS~\cite{Cai_2024_nips_hdrgs}, both of which are trained on identical single-exposure LDR images to ensure fair comparison; and
    \item state-of-the-art single-image HDR reconstruction methods, specifically DrTMO~\cite{endo_2017_tog_drtmo}, SingleHDR~\cite{Liu_2020_CVPR_singlehdr}, and IntrinsicHDR~\cite{dille_2024_eccv_IntrinsicHDR}, integrated with 3D Gaussian Splatting~\cite{kerbl_2023_tog_3Dgaussians}.
\end{enumerate*}
The second category employs pre-trained models on their respective datasets to process input LDR images, followed by training a standard 3DGS on the reconstructed HDR images for novel view synthesis.
From the table we can see that simply combining existing single-image-based HDR reconstruction methods with GS may not work well due to the introduced multi-view inconsistency.
Meanwhile, exiting HDR-NVS methods do not perform well with the single-exposure inputs.
In contrast, our method can reconstruct the HDR scene more accurately and render more photorealistic novel images.
While we note that HDR supervision is often not available in real-world applications, we report the comparisons between our method and the HDR-GS~\cite{Cai_2024_nips_hdrgs} with ground truth HDR images as supervision, revealing the upper bound performance for a reference.

\subsection{Qualitative Results}

\why{
We present a qualitative assessment of HDR novel view synthesis, focusing on visual fidelity across various selected methods. \cref{fig:visual} illustrates comparative results on two synthetic indoor scenes from the HDR-NeRF dataset, evaluating our method alongside established baselines integrated 3DGS.
Visual comparison experiments reveal the results across different methods.  Single-exposure reconstruction approaches (DrTMO~\cite{endo_2017_tog_drtmo}+3DGS, SingleHDR~\cite{Liu_2020_CVPR_singlehdr}+3DGS, IntrinsicHDR~\cite{dille_2024_eccv_IntrinsicHDR}+3DGS) frequently introduce floaters due to limitations in reconstructing HDR from inconsistent multi-view HDR images predicted from LDR inputs. Conversely, multi-exposure methods like HDR-NeRF~\cite{huang_2022_cvpr_hdrnerf} and HDR-GS~\cite{Cai_2024_nips_hdrgs}, when trained on single-exposure data, exhibit unrealistic color tones or blurry regions, attributable to the misalignments between multi-exposures. In contrast, our method, leveraging the proposed \bracketgs and NeEF framework, demonstrates improved coherence and realism, closely aligning with the ground truth by effectively modeling exposure variations in the linear radiance domain and neural fusion. These qualitative results support the robustness of our approach, consistent with the quantitative improvements reported in \cref{tab:quantitative-avg}.
}

\begin{figure}[ht]
\renewcommand{\tabcolsep}{0.8pt}
\renewcommand{\arraystretch}{0.6}

    \centering
    \resizebox{0.9\linewidth}{!}{
        \begin{tabular}{cccc}
            \includegraphics[width=0.24\linewidth]{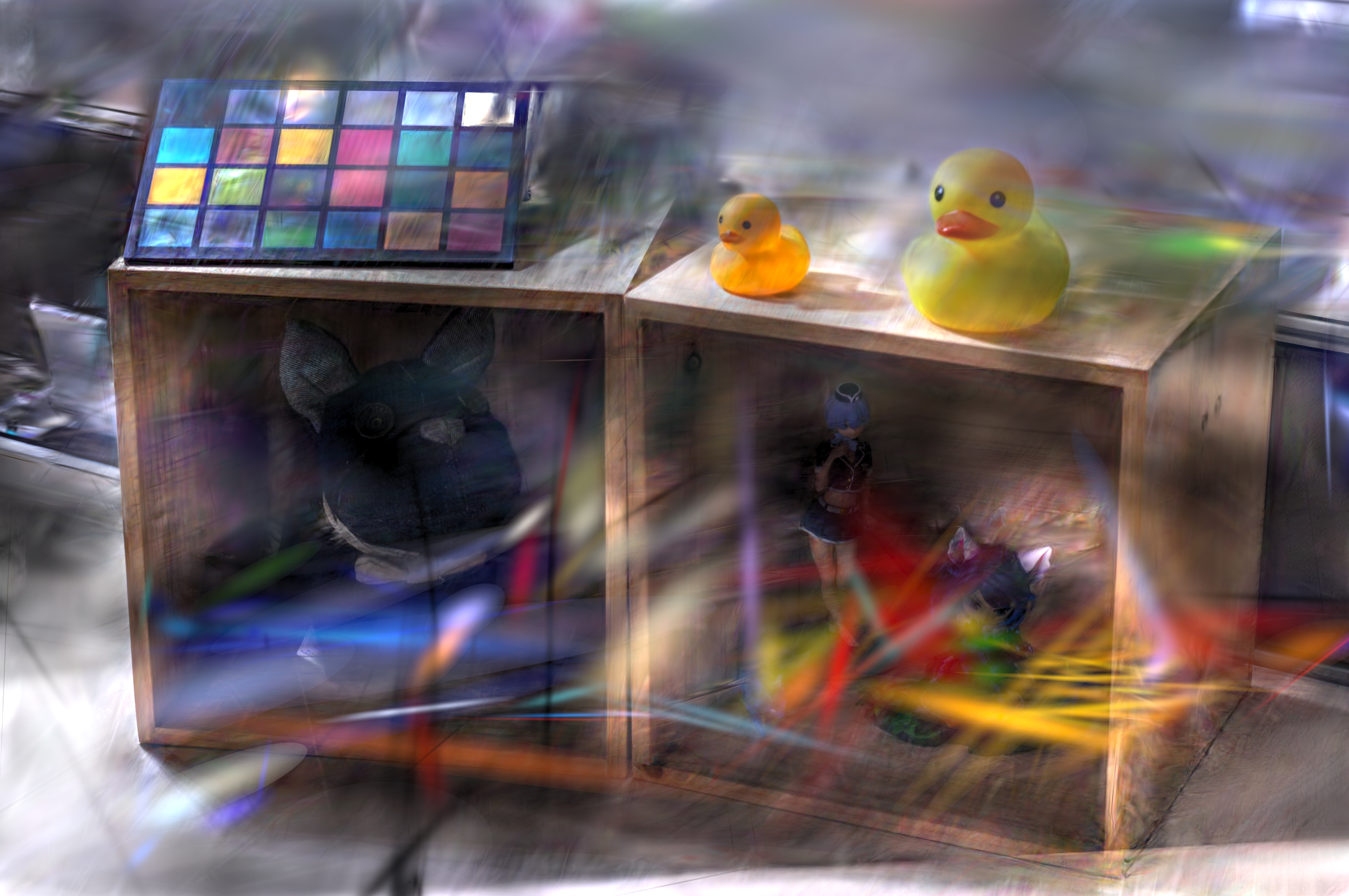} &
            \includegraphics[width=0.24\linewidth]{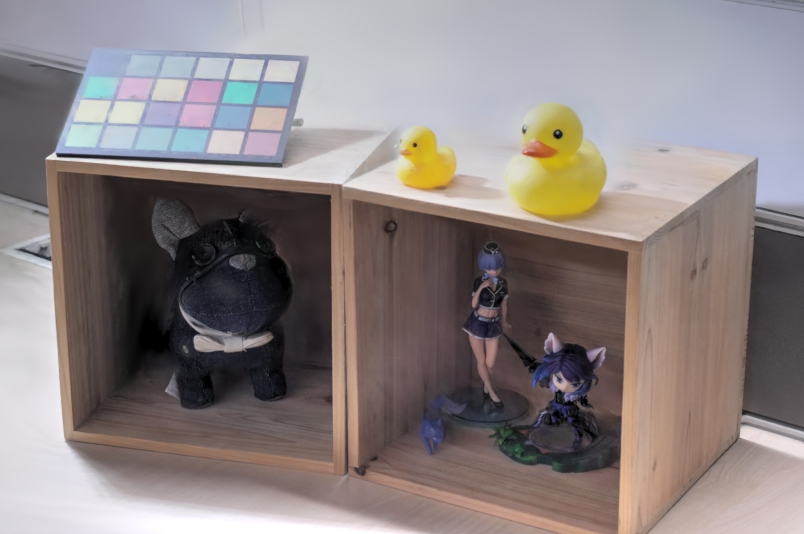} &
            \includegraphics[width=0.24\linewidth]{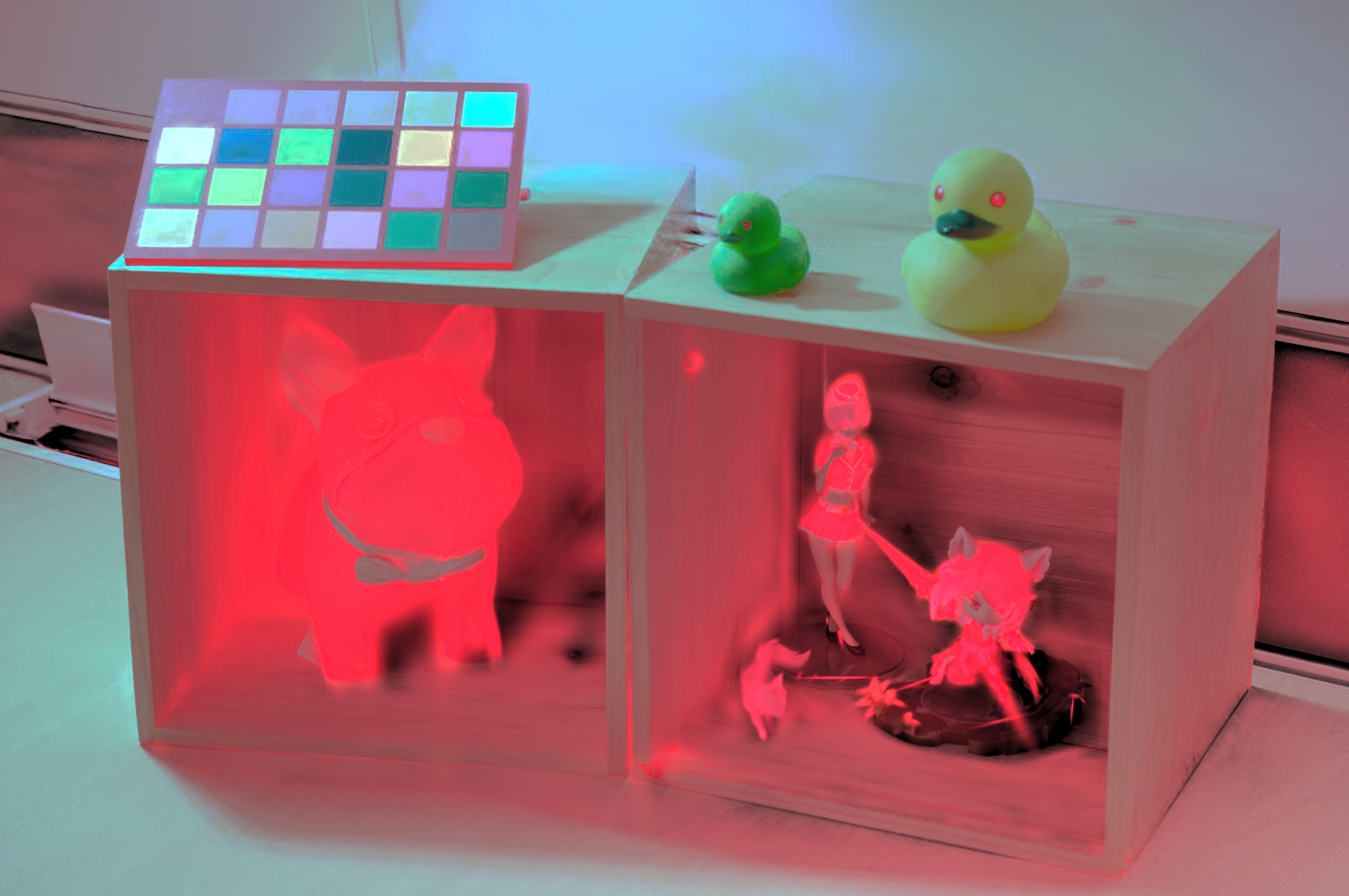} &
            \includegraphics[width=0.24\linewidth]{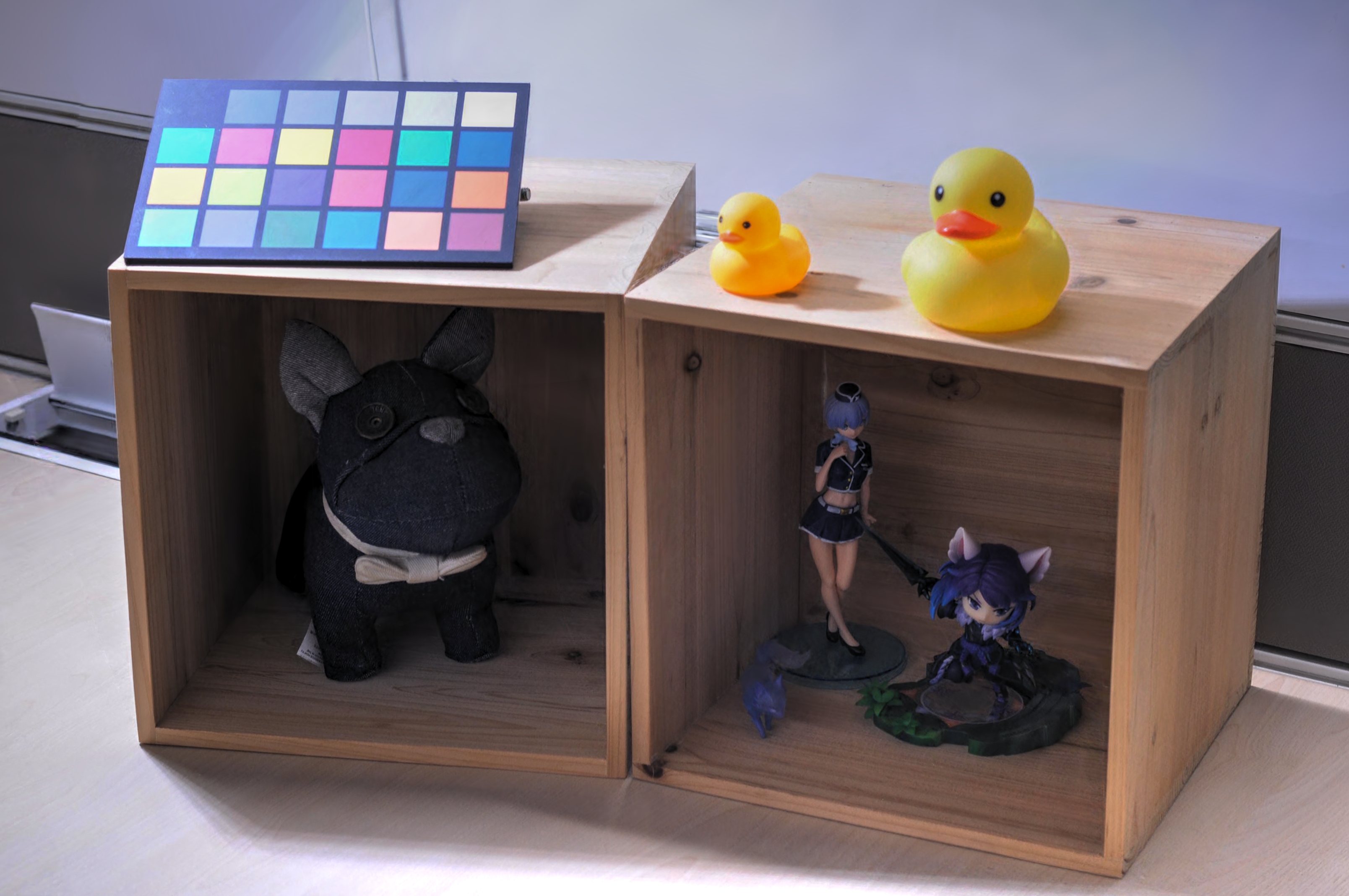} \\

            \scriptsize{SingleHDR+GS} & 
            \scriptsize{IntrinHDR+GS} & 
            \scriptsize{HDRGS} & 
            \scriptsize{Ours} \\

            \includegraphics[width=0.24\linewidth]{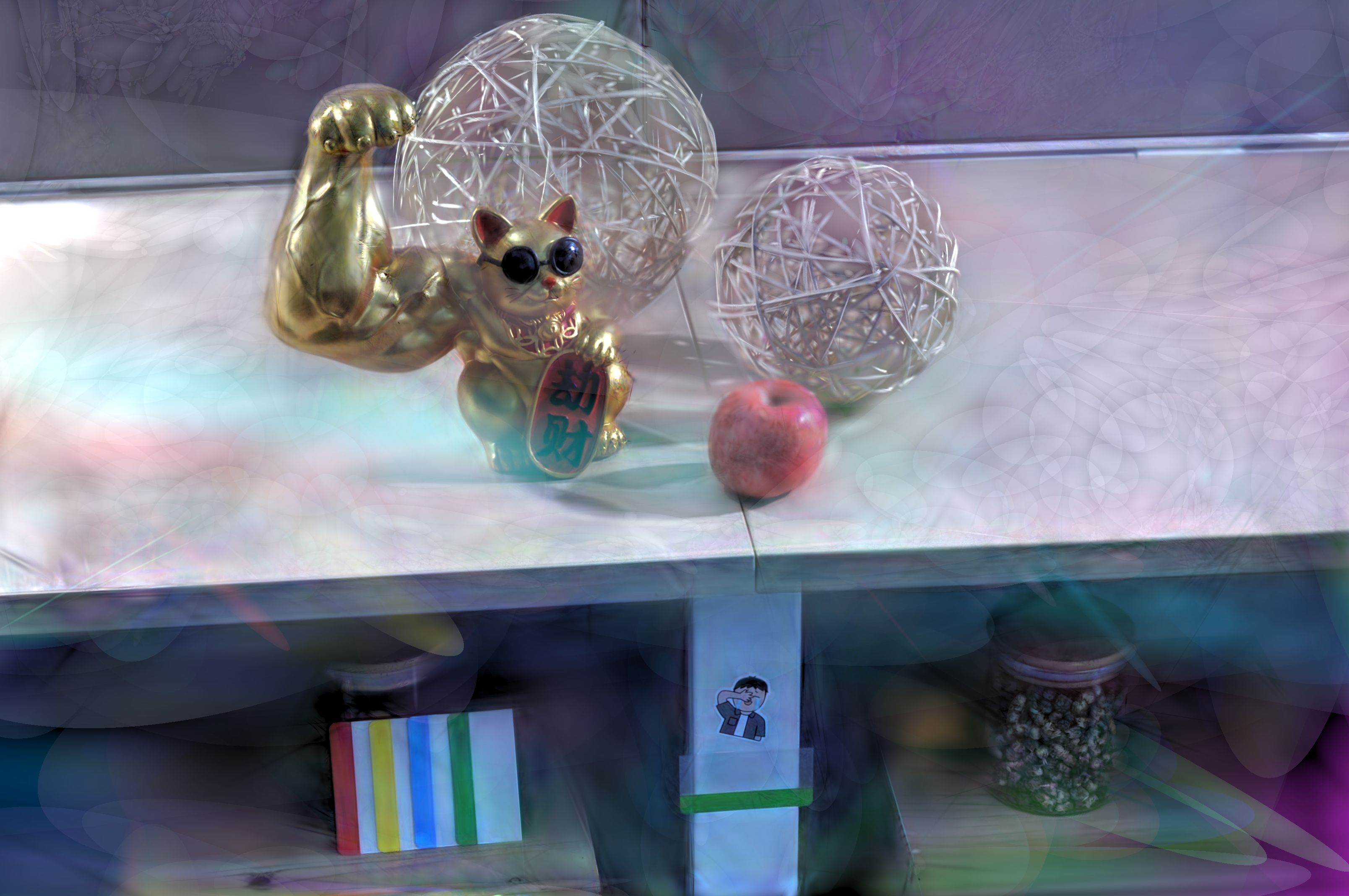} &
            \includegraphics[width=0.24\linewidth]{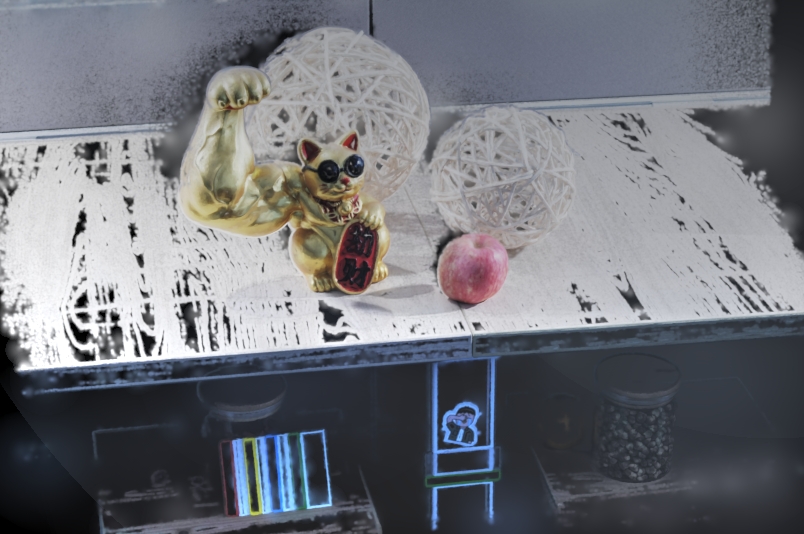} &
            \includegraphics[width=0.24\linewidth]{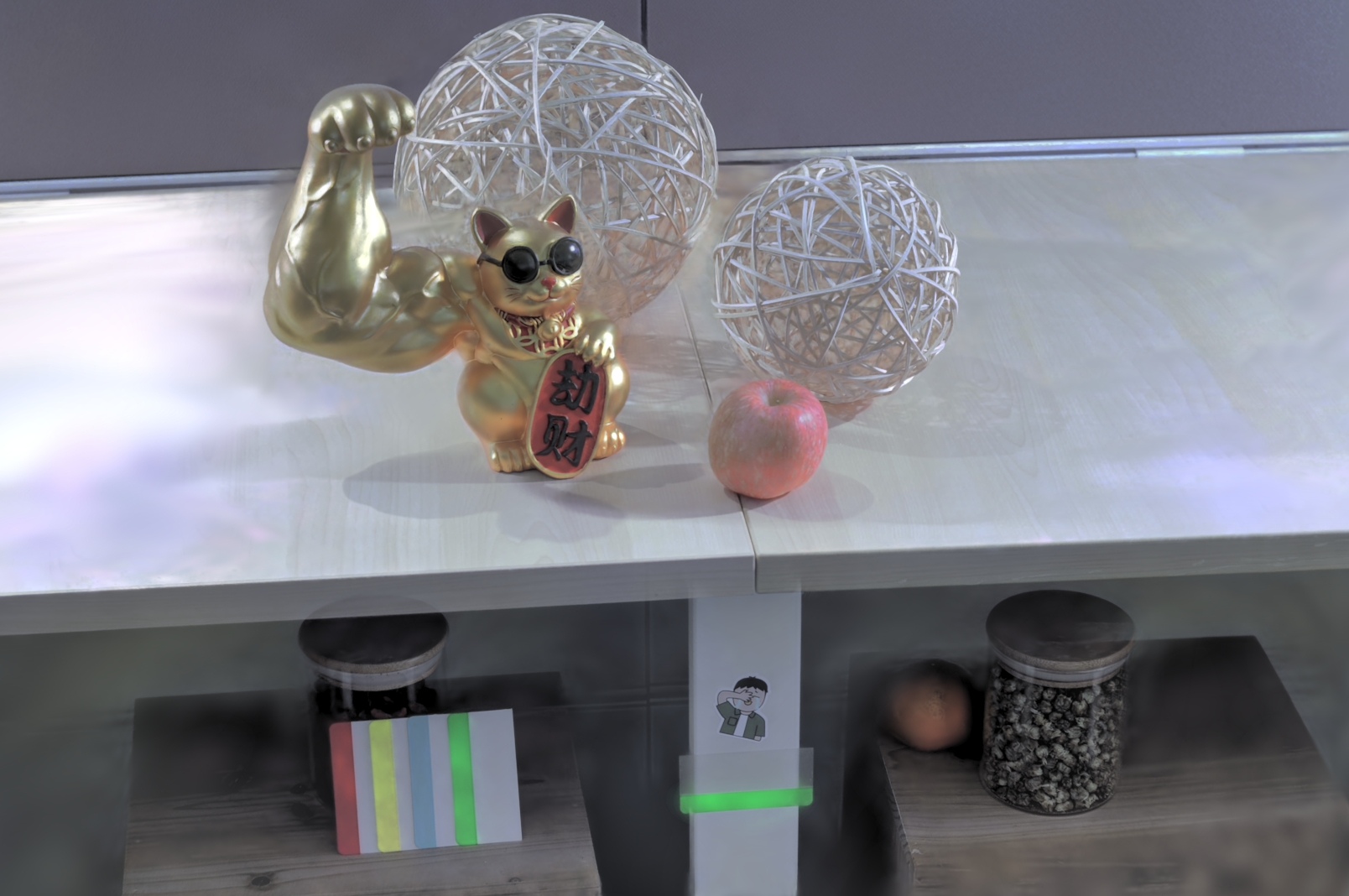} &
            \includegraphics[width=0.24\linewidth]{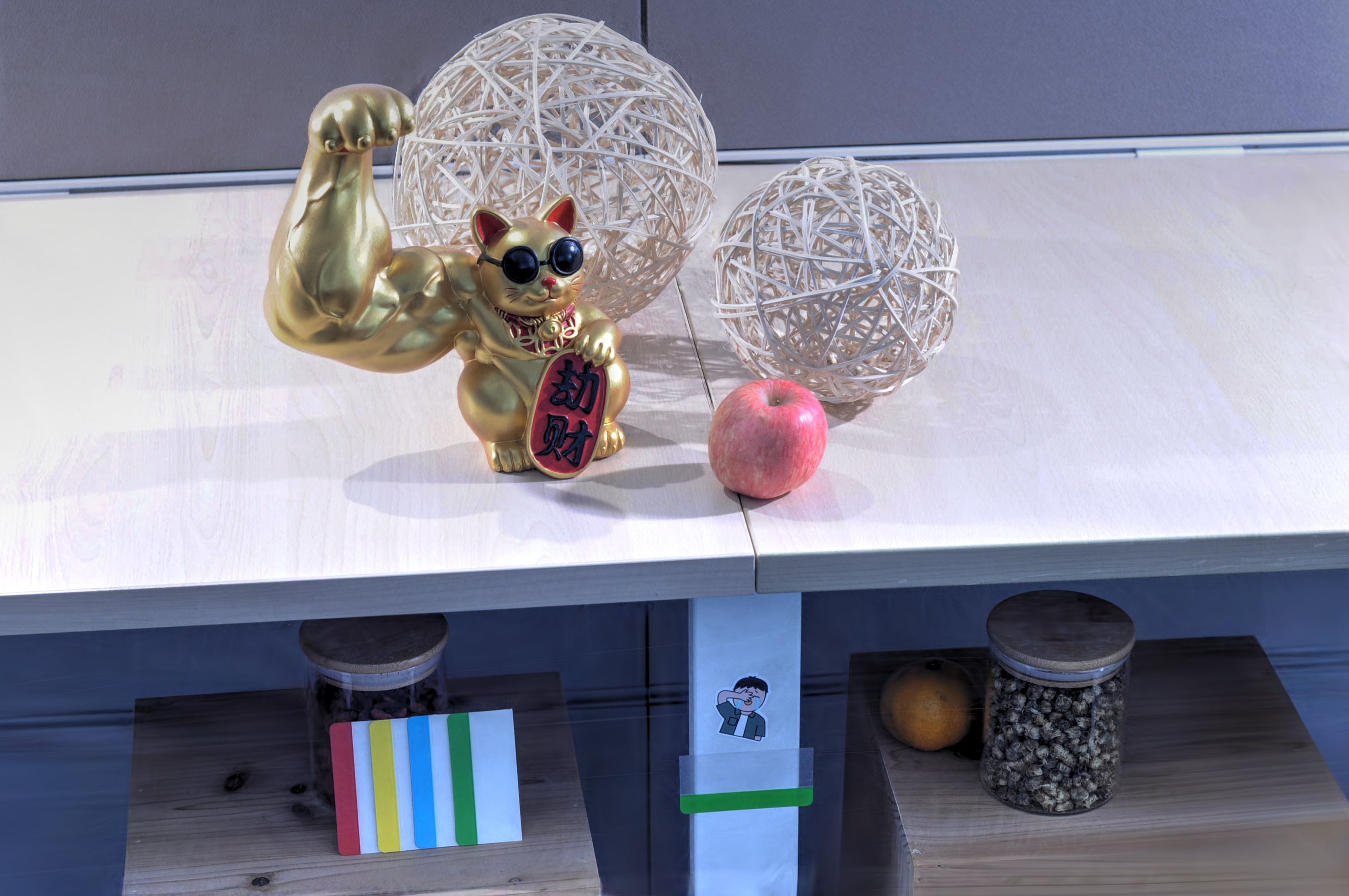} \\

            \scriptsize{SingleHDR+GS} & 
            \scriptsize{IntrinHDR+GS} & 
            \scriptsize{HDRGS} & 
            \scriptsize{Ours} \\
        \end{tabular}
    }
    \vspace{-2mm}
    % \caption{Interior Analysis of \bracketgs.}
    \caption{Visual results on HDR-NeRF~\cite{huang_2022_cvpr_hdrnerf} real scenes where HDR ground truth is not available. Our result produce the most photorealistic results attribute to the effectiveness of bracketed 3DGS, which is able to fuse reliable HDR Gaussian.}
    \vspace{-2mm}
    \label{fig:real}
\end{figure}

\subsection{Ablation Study}

\begin{figure}[ht]
\renewcommand{\tabcolsep}{0.8pt}
\renewcommand{\arraystretch}{0.6}

    \centering
    \resizebox{0.8\linewidth}{!}{
        \begin{tabular}{ccccc}
            \includegraphics[width=0.19\linewidth]{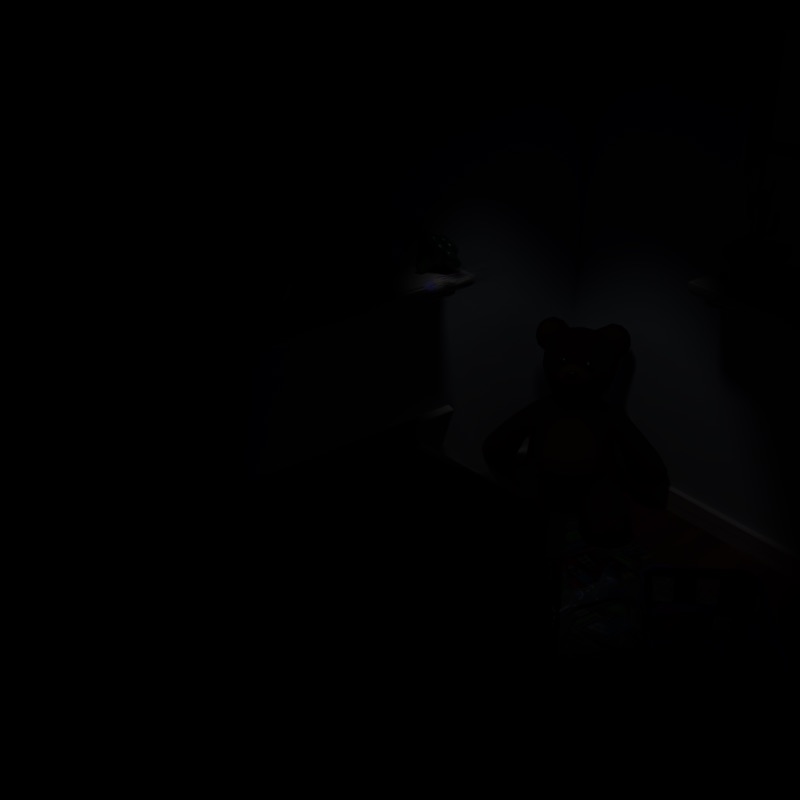} &
            \includegraphics[width=0.19\linewidth]{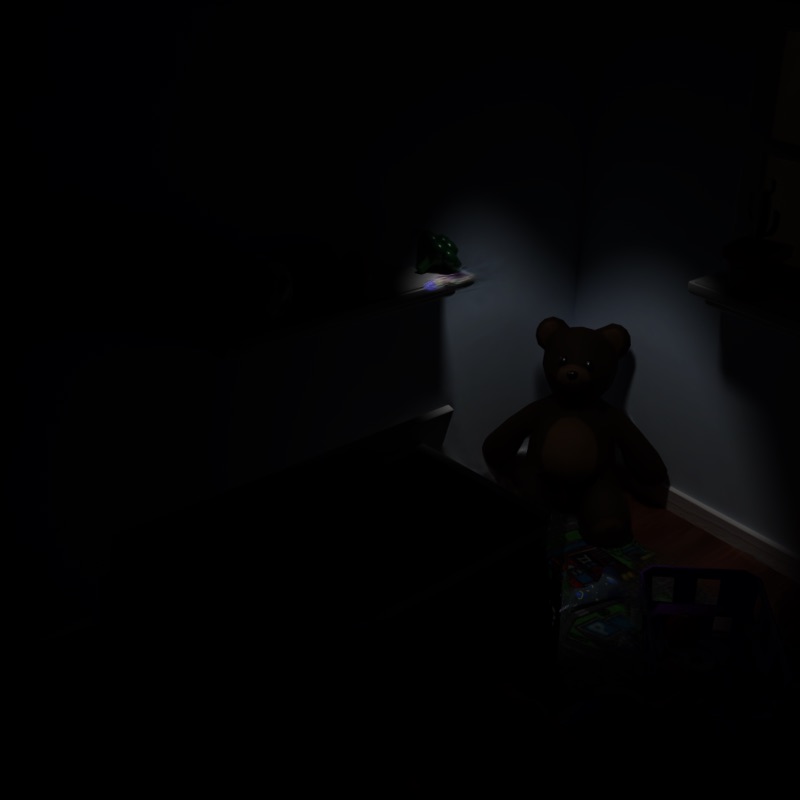} &
            \includegraphics[width=0.19\linewidth]{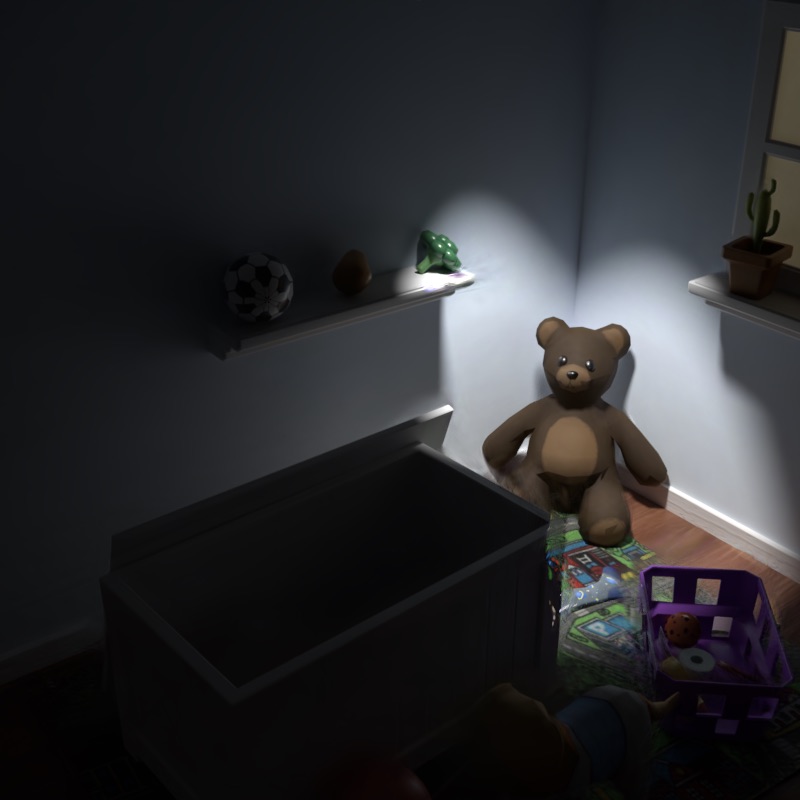} &
            \includegraphics[width=0.19\linewidth]{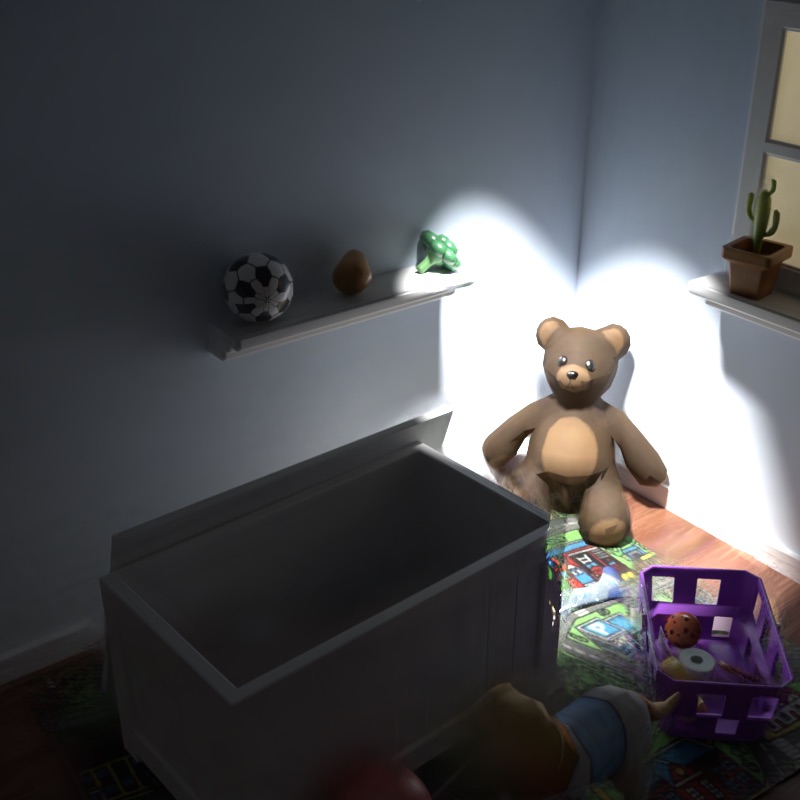} &
            \includegraphics[width=0.19\linewidth]{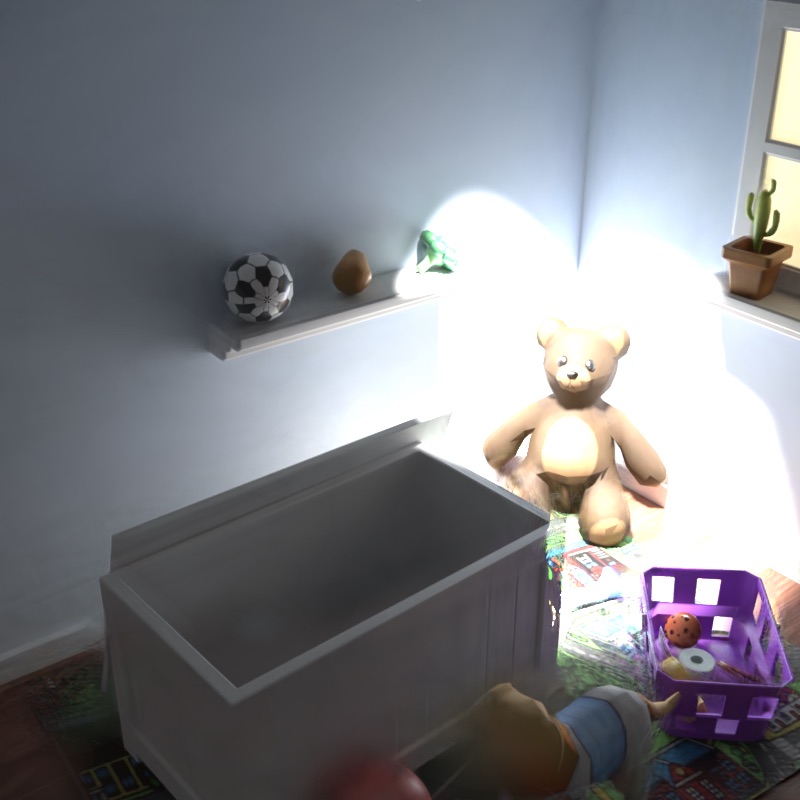} \\

            \includegraphics[width=0.19\linewidth]{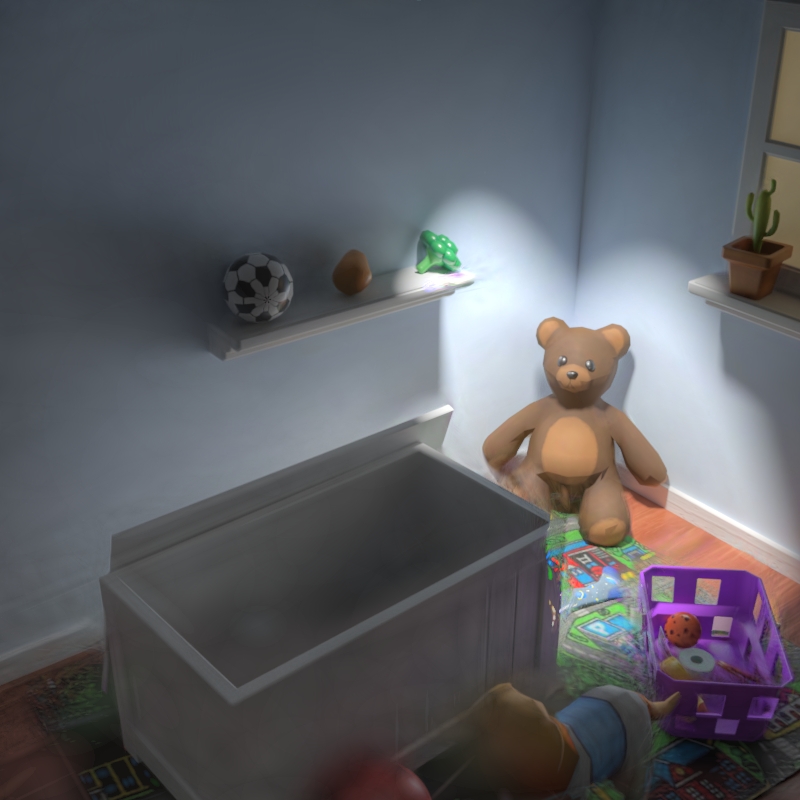} &
            \includegraphics[width=0.19\linewidth]{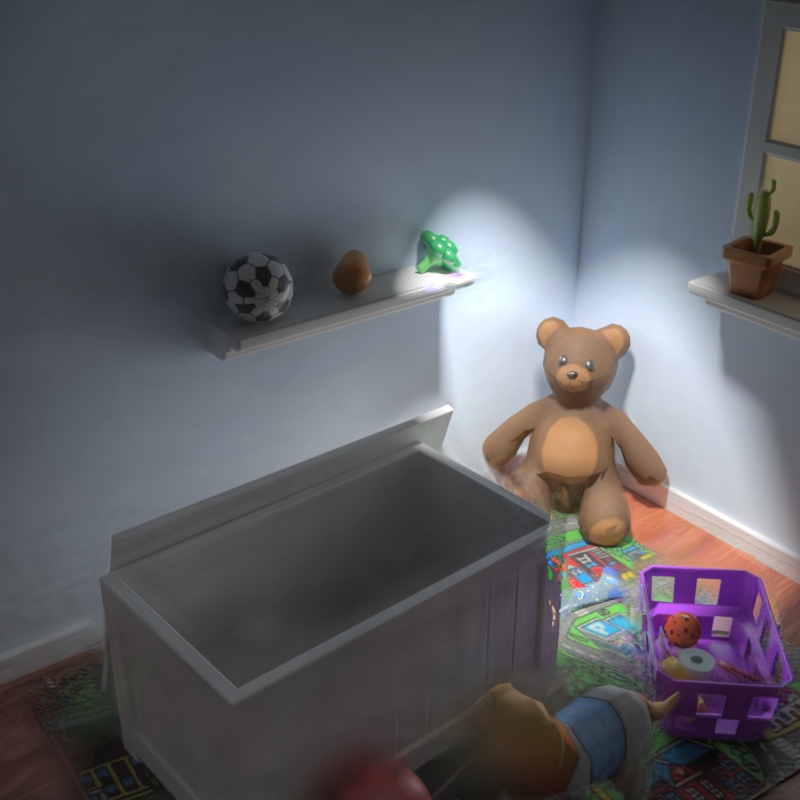} &
            \includegraphics[width=0.19\linewidth]{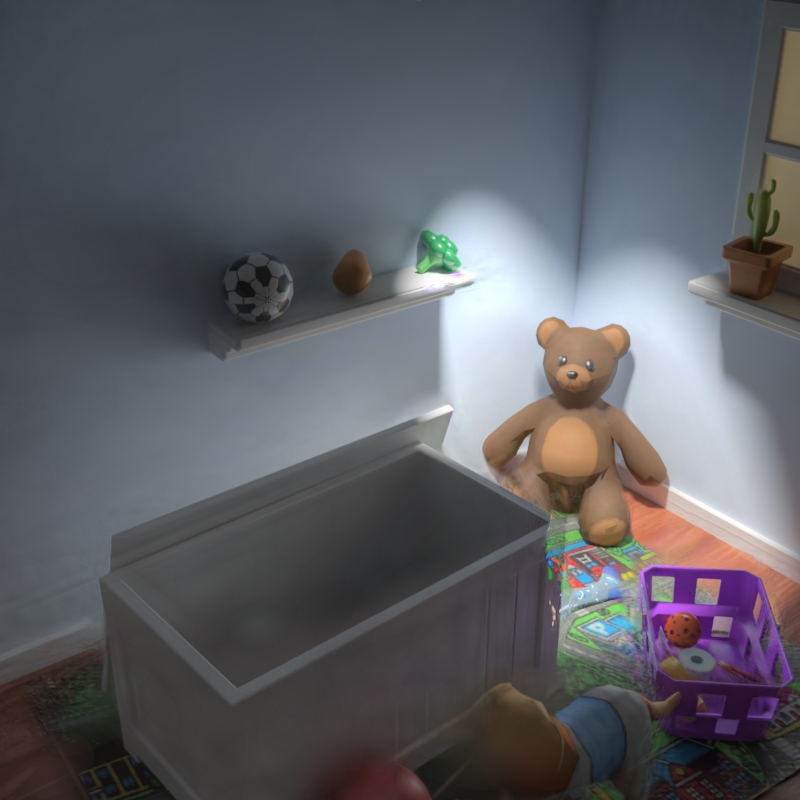} &
            \includegraphics[width=0.19\linewidth]{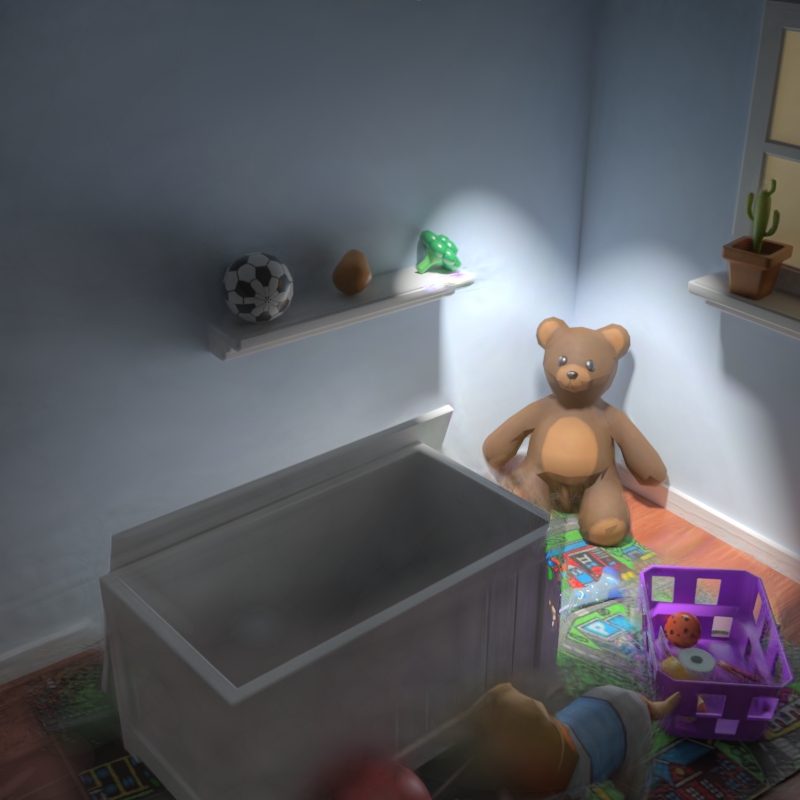} &
            \includegraphics[width=0.19\linewidth]{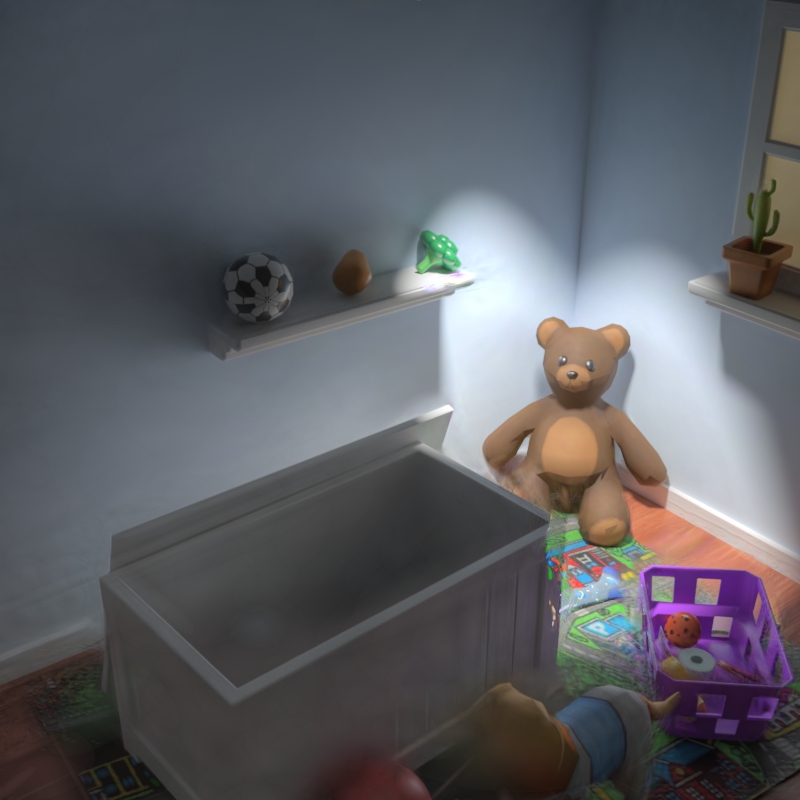} \\

            \scriptsize{$\Delta t=1/8s$} & \scriptsize{$\Delta t=1/4s$} & \scriptsize{$\Delta t=2s$} & \scriptsize{$\Delta t=8s$} & \scriptsize{$\Delta t=16s$} \\

            \includegraphics[width=0.19\linewidth]{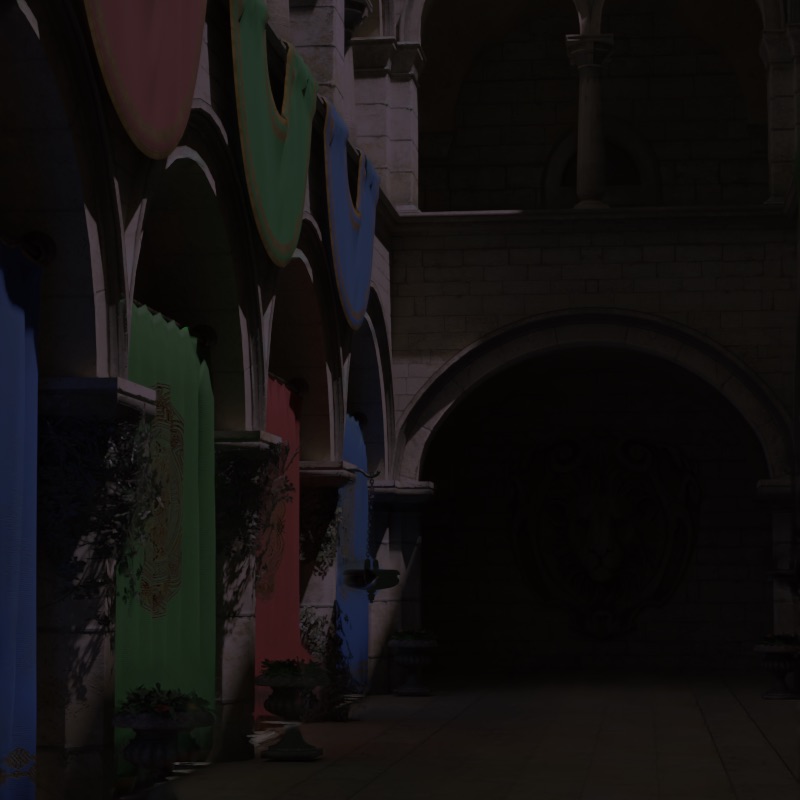} &
            \includegraphics[width=0.19\linewidth]{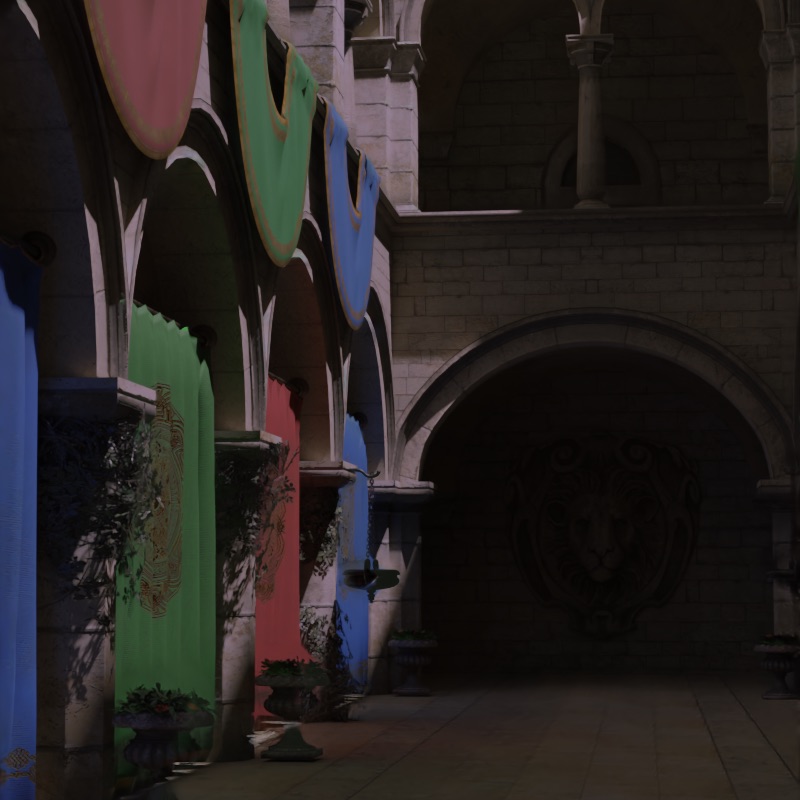} &
            \includegraphics[width=0.19\linewidth]{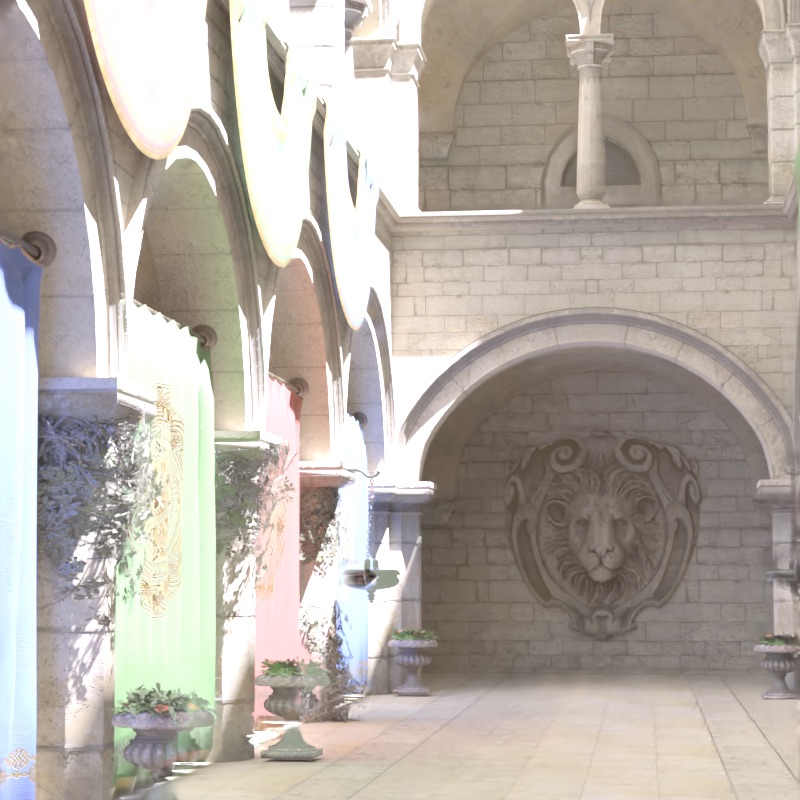} &
            \includegraphics[width=0.19\linewidth]{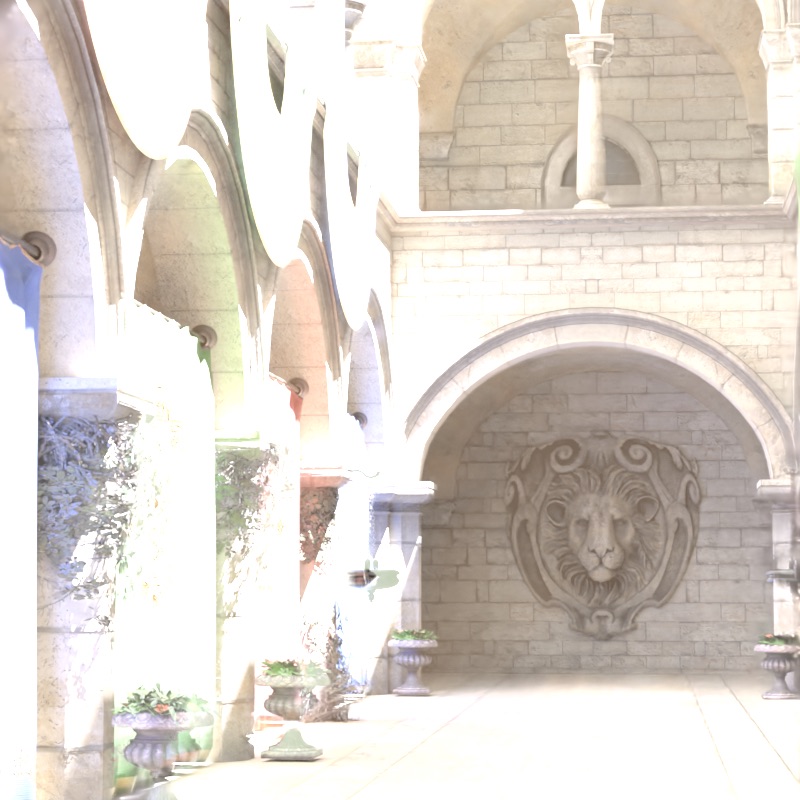} &
            \includegraphics[width=0.19\linewidth]{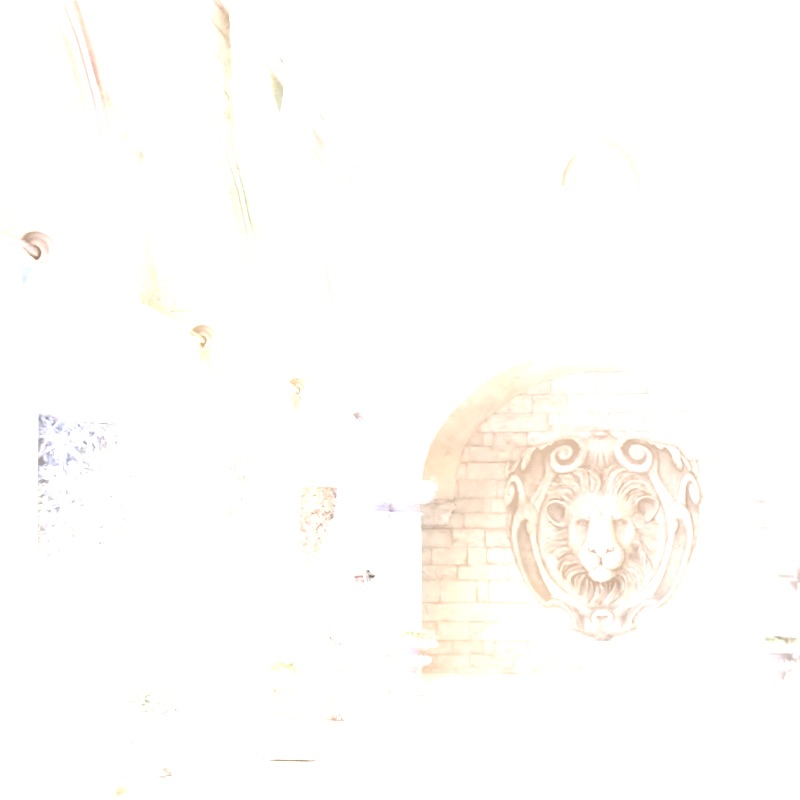} \\

            \includegraphics[width=0.19\linewidth]{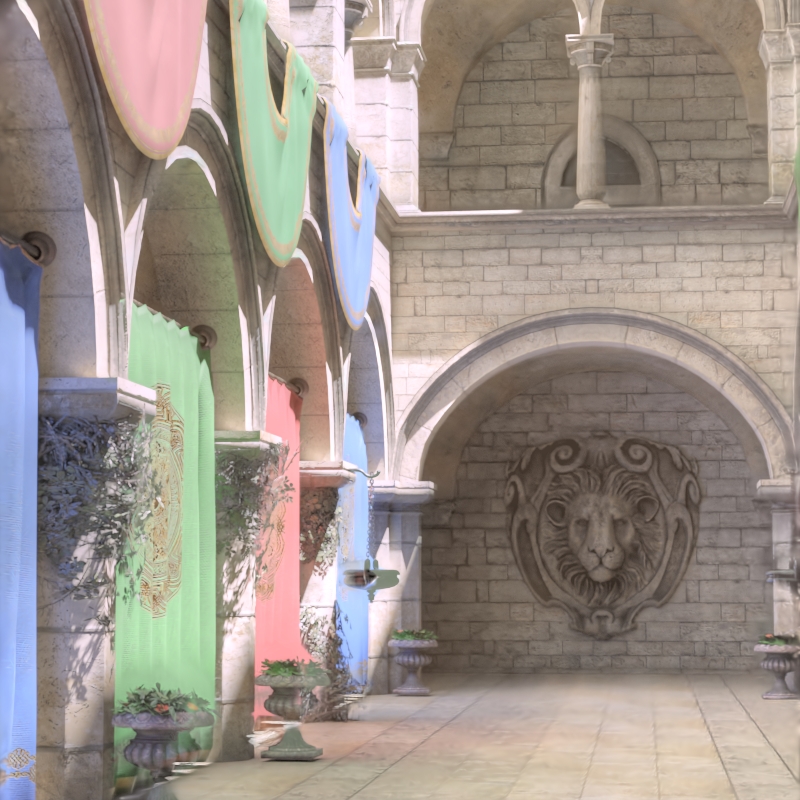} &
            \includegraphics[width=0.19\linewidth]{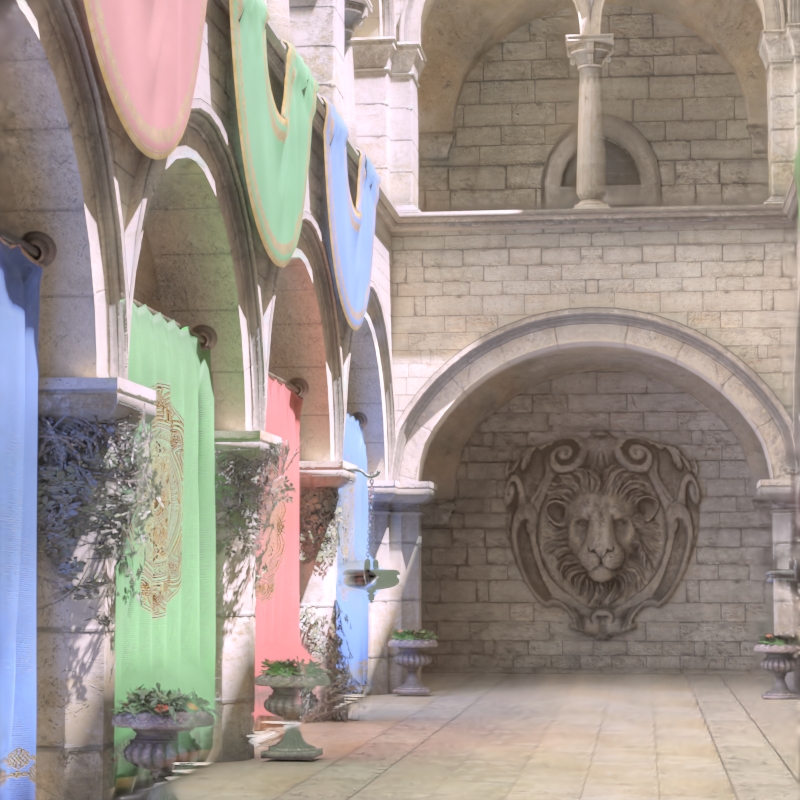} &
            \includegraphics[width=0.19\linewidth]{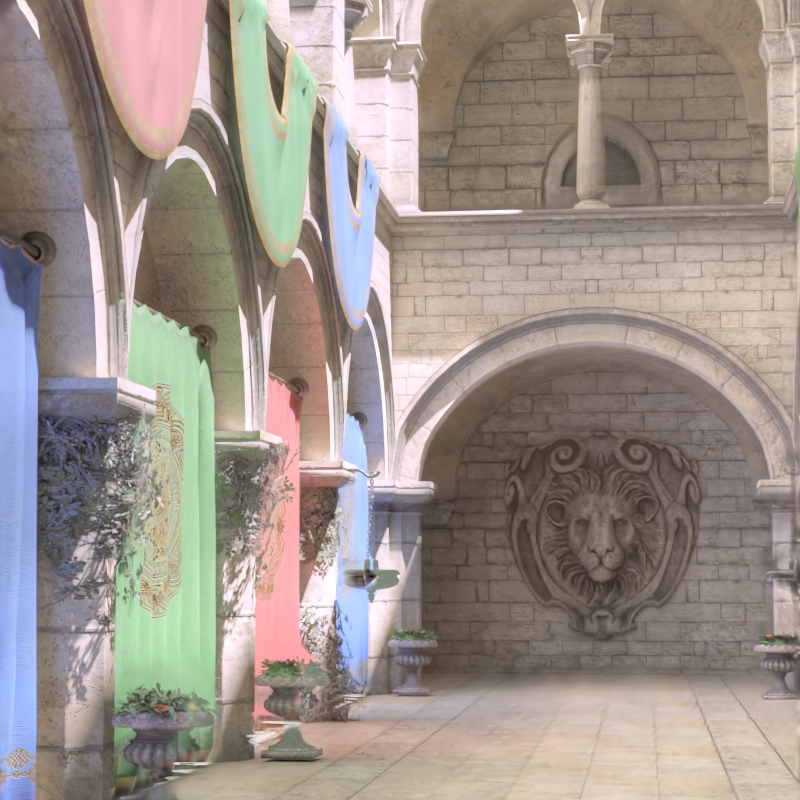} &
            \includegraphics[width=0.19\linewidth]{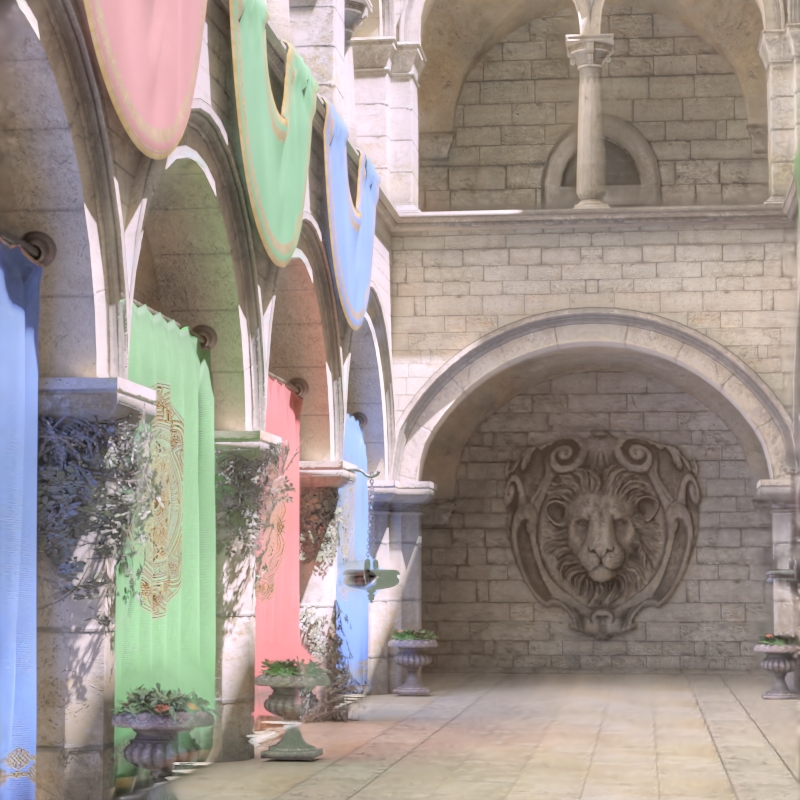} &
            \includegraphics[width=0.19\linewidth]{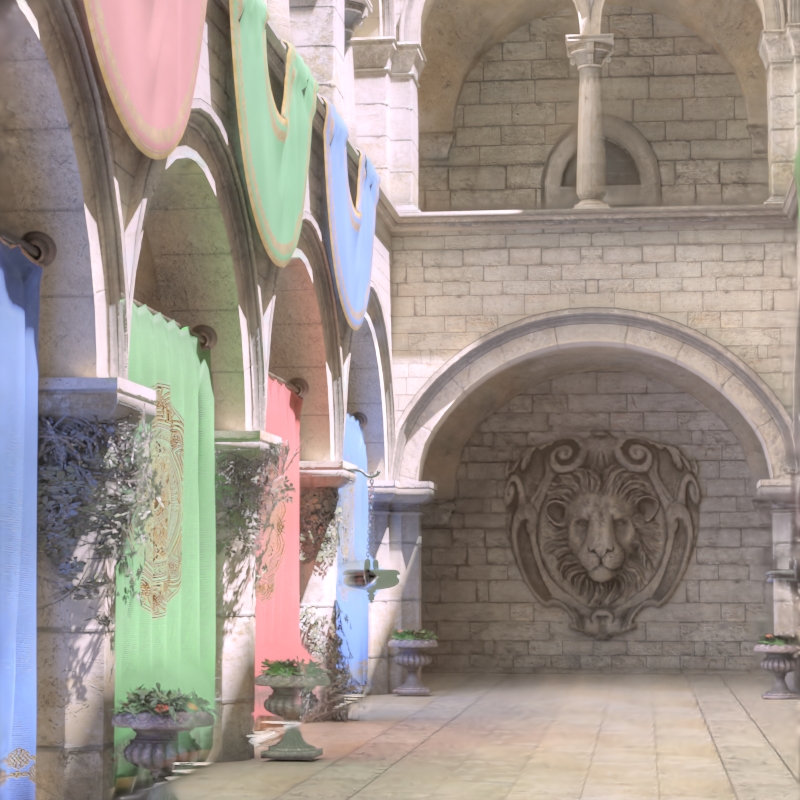} \\

            \scriptsize{$\Delta t=1/8s$} & \scriptsize{$\Delta t=1/4s$} & \scriptsize{$\Delta t=1s$} & \scriptsize{$\Delta t=8s$} & \scriptsize{$\Delta t=16s$} \\
        \end{tabular}
    }
    \vspace{-2mm}
    % \caption{Interior Analysis of \bracketgs.}
    \caption{\why{
    HDR novel views of two scenes are synthesized using bracketed 3D Gaussians with multi-exposure times. Rows 1 and 3 use nonlinear tone-mapping, preserving brightness variations, while Rows 2 and 4 apply linear normalization, ensuring consistency and validating our linear radiance model.
    }
    }
    \vspace{-2mm}
    \label{fig:interior}
\end{figure}

\noindent \textbf{Interior Analysis.}
To validate the \why{effectiveness} of our \bracketgs framework, we render novel views of the bracketed Gaussians in the linear radiance domain, explicitly analyzing exposure-dependent radiance variations to verify the physical plausibility of the multi-exposure modeling strategy.

Given that the linear radiance values $c_i^l$ are inherently proportional to exposure time, applying a linear normalization (\eg, uniformly scaling by maximum number) before tone-mapping will produce identical novel views across different exposures, as the normalization inherently cancels out exposure-dependent scaling.
Conversely, \why{employing a nonlinear normalization before tone-mapping or performing tone-mapping before linear normalization maintains the brightness variations} induced by varying exposures, resulting in distinct intensity distributions across synthesized views.
This behavior validates the physical consistency of our linear radiance modeling across exposures.

\cref{fig:interior} illustrates the comparative visualization of bracketed multi-exposure Gaussians.
Rows 1–2 and Row 3-4 depict HDR novel views synthesized from bracketed Gaussians in the linear radiance domain, processed with distinct tone-mapping strategies:
(1) Row 1 and Row 3 demonstrates non-linear tone-mapping via Apple Core Image filters~\cite{coreimage}, which apply pixel-wise transformations that distort relative brightness and incur non-linearity.
(2) Row 2 and Row 4 shows the results of linearly normalized radiance values followed by Reinhard’s global tone-mapping operator~\cite{Reinhard_2002_tog_photographic} with min-max normalization. %given by:
% Normalization is defined as:
%\begin{align}
%    I = \frac{I - I_\text{min}}{I_\text{max} - I_\text{min}}.
%\end{align}
The non-linear pipeline (Row~1, 3) retains exposure-dependent brightness variations, while linear normalization (Row~2, 4) enforces brightness consistency across exposures by canceling scale differences.
This contrast empirically validates the physical consistency of our linear radiance modeling under exposure variations.

\begin{figure}[t]
\renewcommand{\tabcolsep}{3pt}
\renewcommand{\arraystretch}{0.8}

    \centering
    \resizebox{0.6\linewidth}{!}{
        \begin{tabular}{cc}
            \includegraphics[width=0.49\linewidth]{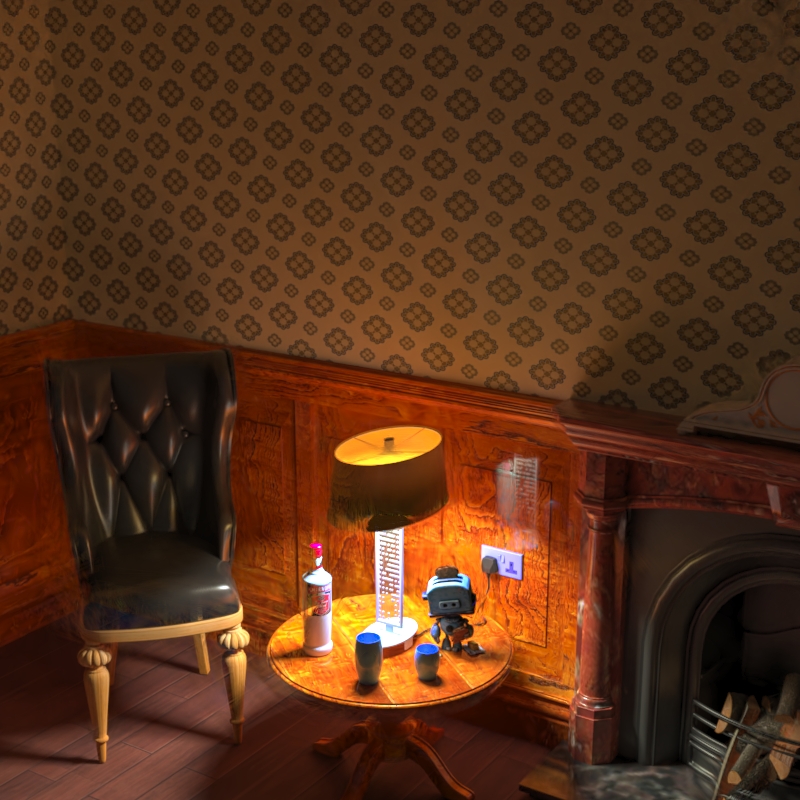} &
            \includegraphics[width=0.49\linewidth]{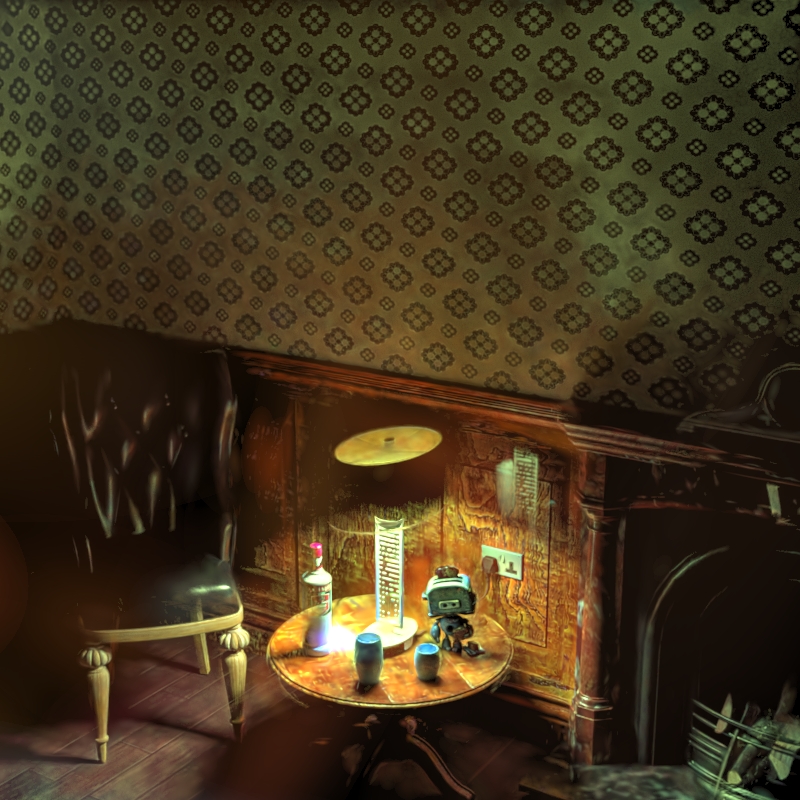} \\

            \Large{w/ $\ell_\text{Linear}$} & 
            \Large{w/o $\ell_\text{Linear}$} \\
        \end{tabular}
    }
    \vspace{-2mm}
    % \caption{Interior Analysis of \bracketgs.}
    \caption{Visual results of ablation on $\ell_\text{Linear}$. The above two figures show the reconstructed HDR novel views under two experiments, with and without our proposed $\ell_\text{Linear}$. Our full model (w/ $\ell_\text{Linear}$) shows better reconstructed HDR color.}
    \vspace{-2mm}
    \label{fig:loss}
\end{figure}

\noindent \textbf{Ablation study.}
To evaluate the effectiveness of the proposed linear loss ($\ell_\text{Linear}$) in ensuring consistent linear radiance estimation, we conduct the ablation experiment, which is performed on the synthetic subsets of the HDR-NeRF dataset~\cite{huang_2022_cvpr_hdrnerf}, and the corresponding results are reported in \cref{fig:loss}. As illustrated in the figure, 
% the results without $\ell_\text{Linear}$ have the problem of \why{???}. In comparison, 
our full model has better visual results, which demonstrates the effectiveness of the proposed loss function. 
% This gain is due to the self-supervised guidance from the linearity prior , which mitigates corrupted radiance estimations observed in unsupervised settings, particularly for real-world scenes lacking ground

\section{Conclusion}
\label{sec:conclusion}

In this work, we have proposed \papertitle, a novel 3DGS-based method to reconstruct and render HDR scenes in novel views given a collection of single-exposure multi-view LDR images.
Our core idea is to estimate and leverage the \bracketgs as a bridge between single-exposure multi-view LDR inputs and the expected 3D HDR scene representation.
To this end, we first learn a base 3D Gaussians from single-exposure LDR inputs.
We then estimate the bracketed 3D Gaussians with identical geometry but different linear colors based on exposure manipulations.
Last, we propose the Differentiable Neural Exposure Fusion (NeEF) to integrate the base and estimated 3D Gaussians into HDR Gaussians for novel view rendering.
Extensive evaluation on standard benchmarks verifying the effectiveness of \papertitle against existing HDR-NVS and baseline methods.

{
    \small
    \bibliographystyle{ieeenat_fullname}
    \bibliography{main}
}

\end{document}